\documentclass[10pt, conference, letterpaper]{IEEEtran}
\usepackage{amssymb} 
\usepackage{url}
\usepackage{mathtools}

\usepackage{tabularx,ragged2e,booktabs}

\usepackage{graphicx}
\DeclareGraphicsExtensions{.pdf,.png,.jpg}
\usepackage{caption,subcaption}
\begin{document}
	\title{MinDelay: Low-latency Forwarding and Caching Algorithms for Information-Centric Networks}
\author{\IEEEauthorblockN{Milad~Mahdian and Edmund~Yeh}
	\IEEEauthorblockA{Department of Electrical and Computer Engineering\\
		Northeastern University\\
		Email: \{mmahdian, eyeh\}@ece.neu.edu}
}
	\newtheorem{defi}{Definition}
	\maketitle
	\begin{abstract}
		We present a new unified framework for minimizing congestion-dependent network cost in information-centric networks by jointly optimizing forwarding and caching strategies.  As caching variables are integer-constrained, the resulting optimization problem is NP-hard.  To make progress, we focus on a relaxed version of the optimization problem, where caching variables are allowed to be real-valued.  We develop necessary optimality conditions for the relaxed problem, and leverage this result to design MinDelay, an adaptive and distributed joint forwarding and caching 
		algorithm, based on the conditional gradient algorithm.  The MinDelay algorithm elegantly yields feasible routing variables and integer caching variables at each iteration, and can be implemented in a distributed manner with low complexity and overhead. 
		Over a wide range of network topologies, simulation results show that MinDelay typically has significantly better delay performance in the low to moderate request rate regions.  Furthermore, the MinDelay and VIP algorithms complement each other in delivering superior delay performance across the entire range of request arrival rates.
		
	\end{abstract}
	
	\section{Introduction}	
	Research on information-centric networking (ICN) architectures over the past few years has brought focus on a number of central network design issues.  One prominent issue is how to jointly design traffic engineering and caching strategies to maximally exploit the bandwidth and storage resources of the network for optimal performance.  While traffic engineering and caching have been investigated separately for many years, their joint optimization within an ICN setting is still an under-explored area.  
	
	There have been many interesting papers on caching strategies within ICN architectures  \cite{Carofiglio-LAC}, \cite{Carofiglio-focal}, \cite{Arald-costaware}, \cite{Thomas-objectoriented}, \cite{Nguyen-congestionprice}, \cite{modeling}, \cite{caching}, \cite{ttlcachingDehgan},\cite{agebasedcachingMing}, \cite{kurose}, \cite{stratis}. 
	When designing and evaluating the effectiveness
	of a cache management scheme
	for such networks, the primary performance metrics
	have been cache hit probability \cite{modeling}, the reduction of the number of hops to retrieve the requested content \cite{caching}, age-based caching \cite{ttlcachingDehgan},\cite{agebasedcachingMing} or content download delay \cite{kurose}. 
	
	Recently, in \cite{stratis}, Ioannidis and Yeh formulate the problem of cost minimization for caching networks with fixed routing and linear link cost functions, and propose an adaptive, distributed caching algorithm which converges to a solution within a (1-1/e) approximation from the optimal.
	

	Similarly, there have been  a number of attempts to enhance the traffic engineering in ICN \cite{giovanna}, \cite{Posch-SAF}, \cite{Detti-modeling}, \cite{mircc}, \cite{Yi-adaptive}. In \cite{giovanna}, Carofiglio et al., formulate the problem of joint multipath congestion control and request forwarding in ICN as an optimization problem. By decomposing the problem into two subproblems of maximizing user throughput and minimizing overall network cost, they develop a receiver-driven window-based Additive-Increase Multiplicative-Decrease (AIMD) congestion control algorithm and a hop-by-hop dynamic request forwarding algorithm which aim to balance the number of pending Interest Packets of each content object (flow) across the outgoing interfaces at each node.  Unfortunately, the work in \cite{giovanna} does not consider caching policies.  
	
	Posch et al. \cite{Posch-SAF} propose a stochastic adaptive forwarding strategy which maximizes the Interest Packet satisfaction ratio in the network. The strategy imitates a self-adjusting water pipe system, where network nodes act as crossings for an incoming flow of water. Each node then intelligently guides Interest Packets along their available paths while circumventing congestion in the system. 
	
	
	In \cite{vip}, Yeh et al., present one of the first unified frameworks for joint forwarding and caching for ICN networks with general topology, in which a virtual control plane operates on the user demand rate for data objects in the network, and an actual plane handles Interest Packets and Data Packets. They develop VIP, a set of distributed and dynamic forwarding and caching algorithms which adaptively maximizes the user demand rate the ICN can satisfy.

	
	
	In this work, we present a new unified framework for minimizing congestion-dependent network cost by jointly choosing node-based forwarding and caching variables, within a quasi-static
	network scenarios where user request statistics vary slowly.   We consider the network cost to be the sum of link costs, expressed as increasing and convex functions of the traffic rate over the links.  When link cost functions are chosen according to an M/M/1 approximation, minimizing the network cost corresponds to minimizing the average request fulfillment delay in the network.  As caching variables are integer-constrained, the resulting joint forwarding and caching (JFC) optimization problem is NP-hard.  To make progress toward an approximate solution, we focus on a relaxed version of the JFC problem (RJFC), where caching variables are allowed to be real-valued.  Using techniques first introduced in~\cite{gallager}, we develop necessary optimality conditions for the RJFC problem.  We then leverage this result to design MinDelay, an adaptive and distributed joint forwarding and caching 
	algorithm for the original JFC problem, based on a version of the conditional gradient, or Frank-Wolfe algorithm.  The MinDelay algorithm elegantly yields feasible routing variables and integer caching variables at each iteration, and can be implemented in a distributed manner with low complexity and overhead. 
	
	Finally, we implement the MinDelay algorithm using our Java-based network simulator, and present extensive experimental results.  We consider three competing schemes, including the VIP algorithm \cite{vip}, which directly competes against MinDelay as a jointly optimized distributed and adaptive forwarding and caching scheme.  Over a wide range of network topologies, simulation results show that while the VIP algorithm performs well in high  request arrival rate regions, MinDelay typically has significantly better delay performance in the low to moderate request rate regions.  Thus, the MinDelay and VIP algorithms complement each other in delivering superior delay performance across the entire range of request arrival rates.

	\section{Network Model}
	Consider a general multi-hop network modeled by a directed and (strongly) connected graph
	$\mathcal{G} = (\mathcal{N} , \mathcal{E})$, where $\mathcal{N}$ and $\mathcal{E}$ are the node and link sets, respectively. A link $(i, j) \in \mathcal{E}$ corresponds to a unidirectional link, with capacity $C_{ij}>0$ ( bits/sec). We assume a content-centric setting, e.g.~\cite{jacobson}, where each node can request any data object from a set of objects $\mathcal{K}$.  A request for a data object consists of a sequence of Interest Packets which request all the data chunks of the object, where the sequence starts with the Interest Packet requesting the starting chunk, and ends with the Interest Packet requesting the ending chunk.  We consider algorithms where the sequence of Interest Packets corresponds to a given object request are forwarded along the same path.
	
	Assume that loop-free routing (topology discovery and data reachability)
	has already been accomplished in the network, so that the Forwarding Interest Base (FIB) tables have been populated for the various data objects.
	Further, we assume symmetric routing, where Data Packets containing the requested data chunks take the same path as their corresponding Interest Packets, in the reverse direction.  Thus, the sequence of Data Packets for a given object request also follow the same path (in reverse).  For simplicity, we do not consider interest suppression, whereby multiple Interest Packets requesting the same  named data chunk are collapsed  into one forwarded Interest Packet.  The algorithm we develop can be extended to include  Interest suppression, by  introducing a virtual  plane in the manner of~\cite{vip}.

	For $k \in \mathcal{K}$, let $src(k)$ be the source node of content object $k$.\footnote{We assume there is one source for each content object for simplicity.  The results generalize easily to the case of multiple source nodes per content object.}  Each node in the network has a local cache of capacity $c_i$ (object units), and can optionally cache Data Packets passing through on the reverse path.  Note that since Data Packets for a given object request follow the same path, all chunks of a data object can be  stored together  at a caching location.  Interest Packets requesting chunks of a given data object can enter the network at any node, and exit the network upon being satisfied by a matching Data Packet at the content source for the object, or at the nodes which decide to cache the object. For simplicity, we assume all data objects have the same size $L$ (bits).  The results of the paper can be extended to the more general case where object sizes differ. 
	
	We focus on quasi-static network scenarios where user request statistics vary slowly. 
	Let $r_i(k)\geq 0$ be the average exogenous rate (in requests/sec)  at which requests for data object $k$ arrive (from outside the network) to node $i$.   Let $t_i(k)$ be the total average arrival rate of object $k$ requests to node $i$.  Thus, $t_i(k)$ includes both the exogenous arrival rate $r_i(k)$ and the endogenous arrival traffic  which is forwarded from other nodes to node $i$. 
	
	Let $ x_i(k) \in \{0,1\}$  be the (integer) caching decision variable for object $k$ at node $i$, where $x_i(k) = 1$ if object $k$ is cached at node $i$ and $x_i(k) =  0$ otherwise.  Thus,
	$t_i(k)x_i(k)$ is the portion of the  total incoming request  rate  for object $k$ which is satisfied from the local cache at node $i$ and $t_i(k)(1-x_i(k))$ is the portion forwarded to neighboring nodes based on the forwarding strategy.  Furthermore, let $\phi_{ij}(k) \in [0,1]$ be the (real-valued) fraction of the traffic $t_i(k)(1-x_i(k))$ forwarded  over link $(i,j)$ by node $i\neq src(k)$.  Thus, $\sum_{j \in \mathcal{O}(i,k)}\phi_{ij}(k) = 1$, where $\mathcal{O}(i,k)$ is the set of neighboring nodes for which node $i$ has a FIB entry for object $k$. Therefore, total average incoming request rate for object $k$ to node $i$ is 
	\begin{equation}
	\label{tik}
	t_i(k) = r_i(k)+\sum_{l \in \mathcal{I}(i,k)} t_l(k)(1-x_l(k)) \phi_{li}(k),
	\end{equation}
	where $\mathcal{I}(i,k)$ is the set of neighboring nodes of $i$ which have FIB entries for node $i$ for object $k$.
	
	Next, let $F_{ij}$ be the Data Packet traffic rate (in bits/sec) corresponding to the total request rate (summed over all data objects) forwarded on link $(i,j) \in \mathcal{E}$:
	\begin{equation}
	\label{fij}
	F_{ij} = \sum_{k \in \mathcal{K}}L\cdot t_i(k)(1-x_i(k))\phi_{ij}(k).
	\end{equation}
	Note that by routing symmetry and per-hop flow balance, the Data Packet traffic of rate $F_{ij}$ actually travels on the reverse link $(j,i)$.
	
	As in \cite{gallager} and \cite{yeh}, we assume the total network cost is the sum of traffic-dependent link costs.  The cost on link $(j,i) \in \mathcal{E}$ is due to the Data Packet traffic of rate $F_{ij}$ generated by the total request rate forwarded on link $(i,j)$, as in~\eqref{fij}.  We therefore denote the cost on link $(j,i)$ as $D_{ij}(F_{ij})$ to reflect this relationship.\footnote{Since Interest Packets are small compared to Data Packets, we do not account for costs associated with the Interest Packet traffic on link $(j,i)$. }  We assume $D_{ij}(F_{ij})$ is increasing and convex in $F_{ij}$. To implicitly impose the link capacity constraint, we assume $D_{ij}(F_{ij}) \rightarrow \infty$ as $F_{ij} \rightarrow C_{ji}^-$ and $D_{ij}(F_{ij}) = \infty$ for $F_{ij} \geq C_{ji}$. As an example, 
	\begin{equation}\label{eq:costfunc}
	D_{ij}(F_{ij}) = \frac{F_{ij}}{C_{ji}-F_{ij}}, \qquad \text{for }0 \leq F_{ij} < C_{ji}.
	\end{equation}
	gives the average number of packets waiting for or under transmission at link $(j,i)$ under an
	$M/M/1$ queuing model \cite{datanetworks}, \cite{kelly}. Summing over all links, the network cost  $\sum_{(i,j)} D_{ij}(F_{ij})$ gives the average total number of packets in the network, which, by Little's Law, is proportional to the average system delay of packets in the network.
	
	\section{Optimization Problem}
	\label{sec:optimization}
	
	We now pose the Joint Forwarding and Caching (JFC) optimization problem in terms of the forwarding variables $(\phi_{ij}(k))_{(i,j) \in \mathcal{E},k \in \mathcal{K}}$ and the caching variables $(x_i(k))_{i \in \mathcal{N},k \in \mathcal{K}}$ as follows:
	\begin{equation}
	\begin{cases}
	\ \min \sum_{(i,j)\in \mathcal{E}} D_{ij}(F_{ij})\\
	\ \text{subject to:}\\
	\ \sum_{j \in \mathcal{O}(i,k)}\phi_{ij}(k) = 1, \qquad \mbox{for~all}~i \in \mathcal{N}, k \in \mathcal{K} \\
	\  \phi_{ij}(k) \geq 0, \qquad \qquad \qquad \mbox{for~all}~(i,j) \in \mathcal{E}, k \in \mathcal{K}\\
	\ \sum_{k \in \mathcal{K}}x_{i}(k) \leq c_i, \qquad \quad\; \mbox{for~all}~i \in \mathcal{N}\\
	\  x_i(k) \in \{0,1\} , \qquad \qquad \quad \mbox{for~all}~i \in \mathcal{N}, k \in \mathcal{K}.
	\end{cases}
	\label{optimization_d}
	\end{equation}
	
	The above mixed-integer optimization problem can be shown to be NP-hard~\cite{femtocaching}.  To make progress toward an approximate solution, we relax the problem by removing the integrality constraint in \eqref{optimization_d}. We formulate the Relaxed JFC (RJFC) problem by replacing the integer caching decision variables $x_i(k) \in \{0,1\}$ by the real-valued variables $\rho_i(k) \in [0,1]$:
	\begin{equation}
	\begin{cases}
	\ \min D  \triangleq \sum_{(i,j)\in \mathcal{E}} D_{ij}(F_{ij})\\
	\ \text{subject to:}\\
	\ \sum_{j \in \mathcal{O}(i,k)}\phi_{ij}(k) = 1, \qquad \mbox{for~all}~i \in \mathcal{N}, k \in \mathcal{K} \\
	\  \phi_{ij}(k) \geq 0, \qquad \qquad \qquad\mbox{for~all}~(i,j) \in \mathcal{E}, k \in \mathcal{K}\\
	\ \sum_{k \in \mathcal{K}}\rho_{i}(k) \leq c_i, \qquad \quad\; \mbox{for~all}~i \in \mathcal{N}\\
	\  0\leq \rho_i(k) \leq 1, \qquad \qquad \quad \mbox{for~all}~i \in \mathcal{N}, k \in \mathcal{K}.
	
	\end{cases}
	\label{optimization}
	\end{equation}
	
	It can be shown that $D$ in \eqref{optimization} is non-convex with respect to (w.r.t.) $(\mathbb{\phi}, \mathbb{\rho})$, where $\mathbb{\phi} \equiv (\phi_{ij}(k))_{(i,j) \in \mathcal{E},k \in \mathcal{K}}$ and the caching variables $\mathbb{\rho}\equiv (x_i(k))_{i \in \mathcal{N},k \in \mathcal{K}}$. In this work, we use the RJFC formulation to develop an adaptive and distributed forwarding and caching algorithm for the JFC problem. 
	
	We proceed by computing the derivatives of $D$ with respect to the forwarding and caching variables, using the technique of~\cite{gallager}. For the forwarding variables, the partial derivatives can be computed as 
	\begin{equation}
	\frac{\partial D}{\partial \phi_{ij}(k)}=(1-\rho_i(k))Lt_i(k)\delta_{ij}(k),
	\label{dif_phi}
	\end{equation}
	where the marginal forwarding cost is
	\begin{equation}
	\delta_{ij}(k) = D'_{ij}(F_{ij}) +\frac{\partial D}{\partial r_j(k)}.
	\label{deltaijk}
	\end{equation}
	
	Note that $\frac{\partial D}{\partial r_j(k)}$ in \eqref{deltaijk} stands for the marginal cost due to a unit increment of object $k$ request traffic at node $j$.  This can be computed recursively by
	$$\frac{\partial D}{\partial r_{j}(k)} =0, \text{ if } j=src(k), $$
	\begin{equation}
	\frac{\partial D}{\partial r_i(k)}=(1-\rho_i(k))L\sum_{j=\mathcal{O}(i,k)}\phi_{ij} (k)\delta_{ij}(k), \text{ if } i\neq src(k).
	\label{dif_r}
	\end{equation}
	
	Finally, we can compute the partial derivatives w.r.t. the (relaxed) caching variables as follows:
	\begin{equation}
	\frac{\partial D}{\partial \rho_i(k)}=-Lt_i(k)\sum_{j=\mathcal{O}(i,k)}\phi_{ij} (k)\delta_{ij}(k).
	\label{dif_rho}
	\end{equation}
	
	The minimization in~\eqref{optimization} is equivalent to minimizing the Lagrangian function 
	\begin{equation}
	\begin{split}
	L(F,\lambda,\mu)=\sum_{(i,j)\in \mathcal{E}}D_{ij}(F_{ij})-\sum_{i,k}\lambda_{ik}\left(\sum_{j}\phi_{ij} (k)-1\right)+\\
	\sum_{i }\mu_i\left(\sum_{k\in \mathcal{K}} \rho_i(k)-c_i\right).
	\end{split}
	\label{lagrangian}
	\end{equation}
	subject to the following constraints:
	\begin{eqnarray*}
		0 \leq \rho_{i}(k) \leq 1, & & \mbox{for~all}~i \in \mathcal{N}, k \in \mathcal{K},\\
		\phi_{ij}(k) \geq 0,  & & \mbox{for~all}~(i,j) \in \mathcal{E}, k \in \mathcal{K}, \\
		\mu_i\geq 0,  & & \mbox{for~all}~i \in \mathcal{N}.
	\end{eqnarray*}
	
	A set of necessary conditions for a local minimum to the RJFC problem can now be derived as
	\begin{equation}
	\frac{\partial D}{\partial \phi_{ij}(k)}\begin{cases}
	\ =\lambda_{ik}, \qquad \text{ if }  \phi_{ij}(k)>0 \\
	\ \geq \lambda_{ik}, \qquad \text{ if }  \phi_{ij}(k)=0
	\end{cases}
	\label{necessary:phi}
	\end{equation}
	\begin{equation}
	\frac{\partial D}{\partial \rho_{i}(k)}\begin{cases}
	\ =-\mu_{i}, \qquad \text{ if }  0<\rho_{i}(k)<1 \\
	\ \geq -\mu_{i}, \qquad \text{ if }  \rho_{i}(k)=0\\
	\ \leq -\mu_{i}, \qquad \text{ if }  \rho_{i}(k)=1
	\end{cases}
	\label{necessary:rho}
	\end{equation}
	with the complementary slackness condition
	\begin{equation}
	\mu_i \left(\sum_{k\in \mathcal{K}} \rho_i(k)-c_i\right) = 0, \mbox{for~all}~i \in \mathcal{N}.
	\label{slackness}
	\end{equation}
	
	The conditions~\eqref{necessary:phi}-\eqref{slackness} are necessary for a local minimum to the RJFC problem, but upon closer examination, it can be seen that they are not sufficient for optimality.  An example from~\cite{gallager} shows a forwarding configuration (without caching)  where~\eqref{necessary:phi} is satisfied at every node,  and yet the operating point is not optimal.  In that example, $t_i(k)=0$ at some node $i$, which leads to \eqref{necessary:phi} being automatically satisfied for node $i$.  This degenerate example applies as well to the joint forwarding and caching setting considered here.  
	
\begin{figure}
		\centering
		\includegraphics[width=.5\textwidth]{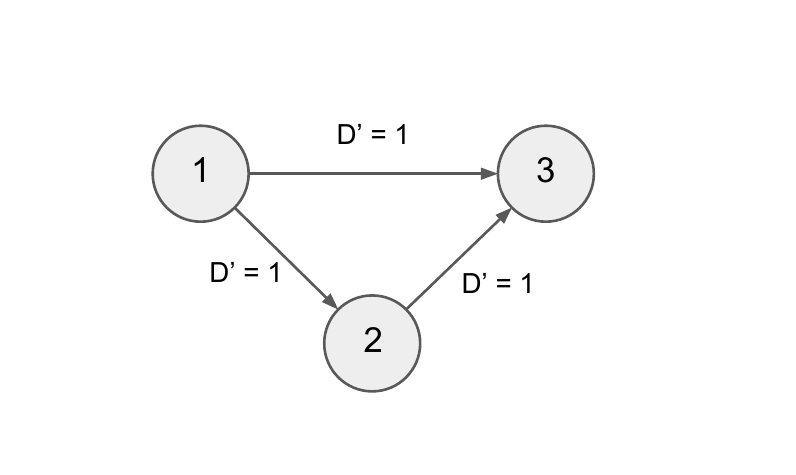}
		\caption{A simple topology.}
		\label{fig:example}
	\end{figure}

	A further issue arises for the joint forwarding and caching setting where $\rho_i(k) = 1$ for some $i$ and $k$.  In this case, the condition in (\ref{necessary:phi}) at node $i$ is automatically  satisfied for every $j \in \mathcal{O}(i,k)$, and yet the operating point need not be optimal.  To illustrate this, consider the simple network shown in Figure \ref{fig:example} with two objects 1 and 2, where $r_1(1)=1$, $r_1(2)=1.5$, $c_1=1$, $c_2=0$ and $src(1)=src(2) =3$. At a given operating point, assume $\rho_1(1) =1$, $ \phi_{12}(1)=1$ and $ \phi_{13}(2)=1$.  Thus, $\rho_1(2) =0$, $\phi_{13}(1)=0$ and $ \phi_{12}(2)=0$. It is easy to see that all the conditions in (\ref{necessary:phi}) and (\ref{necessary:rho}) are satisfied.  However, the current operating point is not optimal. An optimal point is in fact reached when object 2 is cached at node 1, instead. That is, $\rho_1(2) =1$, $ \phi_{13}(1)=\phi_{13}(2)=1$. 
	
	This example, along with the example in~\cite{gallager}, show that when $\rho_i(k) = 1$ or $t_i(k)= 0$, node $i$ still needs to assign forwarding variables for object $k$ in the optimal way.  In the degenerate cases where $\rho_i(k) = 1$ or $t_i(k)= 0$, removing the term 
	$t_i(k)(1-\rho_i(k))$ from~\eqref{necessary:phi} prevents non-optimal forwarding choices.  Furthermore, since the term $t_i(k)(1-\rho_i(k))$ is not a function of $j \in \mathcal{O}(i,k)$, it can also be removed from condition \eqref{necessary:phi} when $t_i(k)(1-\rho_i(k))>0$. We therefore focus on the following modified conditions   
	\begin{equation}
	\delta_{ij}(k) \begin{cases}
	\ =\delta_{i}(k) , \qquad \text{ if }  \phi_{ij}(k)>0 \\
	\ \geq \delta_{i}(k) , \qquad \text{ if }  \phi_{ij}(k)=0.
	\end{cases}
	\label{sufficient:phi}
	\end{equation}
	
	\begin{equation}
	t_i(k)\delta_{i}(k) \begin{cases}
	\ =\mu_{i}, \qquad \text{ if }  0<\rho_{i}(k)<1 \\
	\ \leq \mu_{i}, \qquad \text{ if }  \rho_{i}(k)=0\\
	\ \geq \mu_{i}, \qquad \text{ if }  \rho_{i}(k)=1.
	\end{cases}
	\label{sufficient:rho}
	\end{equation}
	
	where 
	\begin{equation}
	\delta_i(k)=\min_{m\in\mathcal{O}(i,k)}\delta_{im}(k).
	\label{deltaik}	 
	\end{equation}
	
	In (\ref{sufficient:rho}), we used the fact that $\sum_{j=\mathcal{O}(i,k)}\phi_{ij} (k)\delta_{ij}(k) = \delta_i(k)$ if condition (\ref{sufficient:phi}) is satisfied. 
	Condition (\ref{sufficient:rho}) suggests a structured caching policy.  If we sort the data objects in decreasing order with respect to the ``cache scores" $\{t_i(k)\delta_{i}(k)\}$, and cache the top $c_i$ objects, i.e. set $\rho_i(k) = 1$ for the top $c_i$ objects, then condition (\ref{sufficient:rho}) is satisfied. This is indeed the main idea underlying our proposed caching algorithm described in the next section.
	

	\section{Distributed Algorithm: MinDelay}
	\label{sec:algorithm}
	
	The conditions in~\eqref{sufficient:phi}-\eqref{sufficient:rho} give the general structure for a joint forwarding and caching algorithm for solving the RJFC problem. For forwarding, each node $i$ must decrease those forwarding variables $\phi_{ij}(k)$ for which the marginal forwarding cost $\delta_{ij}(k)$ is large, and increase those for which it is small.  For caching, node $i$ must increase the caching variables $\rho_i(k)$ for which the cache score $ t_i(k)\delta_{i}(k)$ is large and decrease those for which it is small.  To  describe the joint forwarding and caching algorithm, we first describe a protocol for calculating the marginal costs, and then  describe an algorithm for updating the forwarding and caching variables. 
	
	
	Note that each node $i$ can estimate, as a time average, the link traffic rate $F_{ij}$ for each outgoing link $(i,j)$. This can be done by either measuring the rate of received Data Packets on each of the corresponding incoming links $(j,i)$, or by measuring the request rate of Interest Packets forwarded on the outgoing links $(i,j)$.  Thus, given a functional form for $D_{ij}(.)$, node $i$ can compute $D'_{ij}(F_{ij})$. 
	
	Assuming a loop-free routing graph on the network, one has a well-defined partial ordering where a node $m$ is {\em downstream} from node $i$ with respect to object $k$ if there exists a routing path from $m$ to $src(k)$ through $i$. A node $i$ is {\em upstream} from node $m$ with respect to $k$ if $m$ is downstream from $i$ with respect to $k$.  
	
	To update the marginal forwarding costs, the nodes use the following protocol.  Each node $i$ waits until it has received the value ${\partial D}/{\partial r_j(k)}$ from each of its upstream neighbors with respect to object $k$ (with the convention ${\partial D}/{\partial r_{src(k)}(k)} = 0$).  Node $i$ then calculates ${\partial D}/{\partial r_i(k)}$ according to~\eqref{dif_r} and broadcasts this to all of its downstream neighbors with respect to $k$.  The information propagation can be done by either piggybacking on the Data Packets of the corresponding object, or by broadcasting a single message regularly to update the marginal forwarding costs of all the content objects at once.
	
	Having described the protocol for calculating marginal costs, we now specify the algorithm for updating the forwarding and caching variables.  Our algorithm is based on the conditional gradient or Frank-Wolfe algorithm~\cite{nlp}.  Let 
	$${\Phi}^n= \begin{bmatrix}
	\left(\phi^{n}_{ij}(k) \right)_{i\in \mathcal{N}, k\in \mathcal{K},j\in \mathcal{O}(i,k)} \\
	\left(\rho^{n}_{i}(k) \right)_{ i\in \mathcal{N}, k\in \mathcal{K}}
	\end{bmatrix}$$
	be the vector of forwarding and caching variables at iteration $n$. Then, the conditional gradient method is given by
	\begin{equation}
	\Phi^{n+1} = \Phi^{n} + a^n(\bar{\Phi}^{n} -\Phi^{n}), \label{conditional}
	\end{equation}
	where $a^n \in (0,1]$ is a positive stepsize, and $\bar{\Phi}^{n}$ is the solution of the direction finding subproblem
	\begin{equation}
	\bar{\Phi}^{n} \in \arg \min_{\Phi \in F} \triangledown D(\Phi^{n} )' (\Phi -\Phi ^n). \label{opt_direction_gen}
	\end{equation}
	Here, $\triangledown D(\Phi^{n} )$ is the gradient of the objective function with respect to the forwarding and caching variables, evaluated at $\Phi^{n}$.  The set $F$ is the set of forwarding and caching variables $\Phi$ satisfying the constraints in (\ref{optimization}), seen to be a bounded polyhedron.  
	
	The idea behind the conditional gradient algorithm is to iteratively find a descent direction by finding the feasible direction $\bar{\Phi}^{n} -\Phi^{n}$ at a point $\Phi^{n}$, where $\bar{\Phi}^{n}$ is a point of $F$ that lies furthest along the negative gradient direction $-\triangledown D(\Phi^{n})$~\cite{nlp}.
	
	In applying the conditional gradient algorithm,  we encounter the same problem regarding degenerate cases as seen in Section~\ref{sec:optimization} with respect to optimality conditions. Note that when $t_i(k)(1-\rho_i(k))=0$, the $\frac{\partial D}{\partial \phi_{ij}(k)}$ component of $\triangledown D(\Phi^{n})$ is zero, and thus provides no useful information for the optimization in~\eqref{opt_direction_gen} regarding the choice of $\bar{\Phi}^{n}$. On the other hand, when $t_i(k)(1-\rho_i(k))>0$, eliminating this term from $\frac{\partial D}{\partial \phi_{ij}(k)}$ in~\eqref{opt_direction_gen} does not change the choice of $\bar{\Phi}^{n}$, since $t_i(k)(1-\rho_i(k))>0$ is not a function of $j \in \mathcal{O}(i,k)$.  Motivated by this observation, we define 
	\begin{equation}
	G(\Phi^{n} ) \triangleq   \begin{bmatrix}
	\left(\delta^{n}_{ij}(k) \right)_{i\in \mathcal{N}, k\in \mathcal{K},j\in \mathcal{O}(i,k)} \\
	\left(-t^n_i(k)\sum_{j=\mathcal{O}(i,k)}\phi^n_{ij} (k)\delta^n_{ij}(k)\right)_{ i\in \mathcal{N}, k\in \mathcal{K}}
	\end{bmatrix}, \label{eq:gradD}
	\end{equation}
	where $\delta^n_{ij}(k)$ and $t^n_i(k)$ are the marginal forwarding costs and total request arrival rates, respectively, evaluated at $\Phi^{n}$. 
	
	We consider a modified conditional gradient algorithm where the direction finding subproblem is given by
	\begin{equation}
	\bar{\Phi}^{n} \in \arg \min_{\Phi \in F} G(\Phi^{n} )' (\Phi -\Phi ^n). \label{opt_direction}
	\end{equation}
	It can easily be seen that (\ref{opt_direction}) is separable into two subproblems. 
	
	The subproblem for $(\phi_{ij}(k))$ is given by
	\begin{equation}
	\begin{cases}
	\ \min  \sum_{(i,k)} \sum_{j\in \mathcal{O}(i,k)} \delta^n_{ij}(k)(\phi_{ij}(k)-\phi_{ij}^n(k))\\
	\ \text{subject to:}\\
	\ \sum_{j \in \mathcal{O}(i,k)}\phi_{ij}(k) = 1, \qquad \mbox{for~all}~i \in \mathcal{N}, k \in \mathcal{K} \\
	\  \phi_{ij}(k) \geq 0, \qquad \mbox{for~all}~i \in \mathcal{N}, k \in \mathcal{K}, j\in \mathcal{O}(i,k).\\
	\end{cases}
	\label{opt_phi}
	\end{equation}
	where 
	\begin{equation}
	\delta^n_{ij}(k) = D'_{ij}(F^n_{ij}) +\frac{\partial D}{\partial r^n_j(k)}.
	\label{deltanijk}
	\end{equation}
	It is straightforward to verify that a solution $\bar{\phi}^n_i(k) = (\bar{\phi}^n_{ij}(k))_{j\in\mathcal{O}(i,k)}$ to \eqref{opt_phi} has all coordinates equal to zero except for one coordinate, say $\bar{\phi}^n_{im}(k)$, which is equal to 1, where 
	\begin{equation}
	m \in \arg \min_{j\in\mathcal{O}(i,k)} \delta^n_{ij}(k). \label{choice_phi}
	\end{equation}
	corresponds to an outgoing interface with the minimal marginal forwarding cost.  Thus, the update equation for the forwarding variables is: for all $i \in \mathcal{N}$, 
	\begin{equation}
	\phi_{ij}^{n+1}(k) = (1-a^{n})\phi_{ij}^{n}(k) +a^{n}\bar{\phi}_{ij}^{n}(k) , \forall k \in \mathcal{K}, j\in \mathcal{O}(i,k). \label{update_phi}
	\end{equation}
	
	The subproblem for $(\rho_i(k))$ is equivalent to
	\begin{equation}
	\begin{cases}
	\ \max   \sum_{(i,k)}  \omega_i^n(k)(\rho_{i}(k)-\rho_{i}^n(k))\\
	\ \text{subject to:}\\
	\ \sum_{k \in \mathcal{K}}\rho_{i}(k) \leq c_i, \qquad \mbox{for~all}~i \in \mathcal{N}\\
	\ 0 \leq \rho_{i}(k) \leq 1, \qquad \mbox{for~all}~i \in \mathcal{N}, k \in \mathcal{K}.\\
	\end{cases}
	\label{opt_rho}
	\end{equation}
	where $\omega_i^n(k) =t_i^n(k)\left(\sum_{j\in \mathcal{O}(i,k)}\phi^n_{ij}(k) \delta^n_{ij}(k)\right)$. The subproblem (\ref{opt_rho}) is a max-weighted matching problem which has an integer solution. For node $i$, let $\omega_i^n(k_1) \geq \omega_i^n(k_2)\geq \ldots \geq \omega_i^n(k_{|\mathcal{K}|})$ be a re-ordering of the $\omega_i^n(k)$'s in decreasing order.  A solution $\bar{\rho}_i^n$ to \eqref{opt_rho} has 
	$\bar{\rho}_i^n(k) = 1$ for $k \in \{k_1,k_2,\ldots,k_{c_i}\}$, and $\bar{\rho}_i^n(k) = 0$ otherwise.  That is, $\bar{\rho}_i^n(k) = 1$ for the $c_i$ objects with the largest $\omega_i^n(k)$ values, and $\bar{\rho}_i^n(k) = 0$ otherwise.  The update equation for the caching variables is: for all $i \in \mathcal{N}$, 
	\begin{equation}
	\rho_{i}^{n+1}(k) = (1-a^{n})\rho_{i}^{n}(k) + a^{n}\bar{\rho}_{i}^{n}(k) , \mbox{for~all}~k \in \mathcal{K}. \label{update_rho}
	\end{equation}

	\begin{figure*}
		\begin{subfigure}[t]{0.4\textwidth}
			\centering
			\includegraphics[scale=0.35]{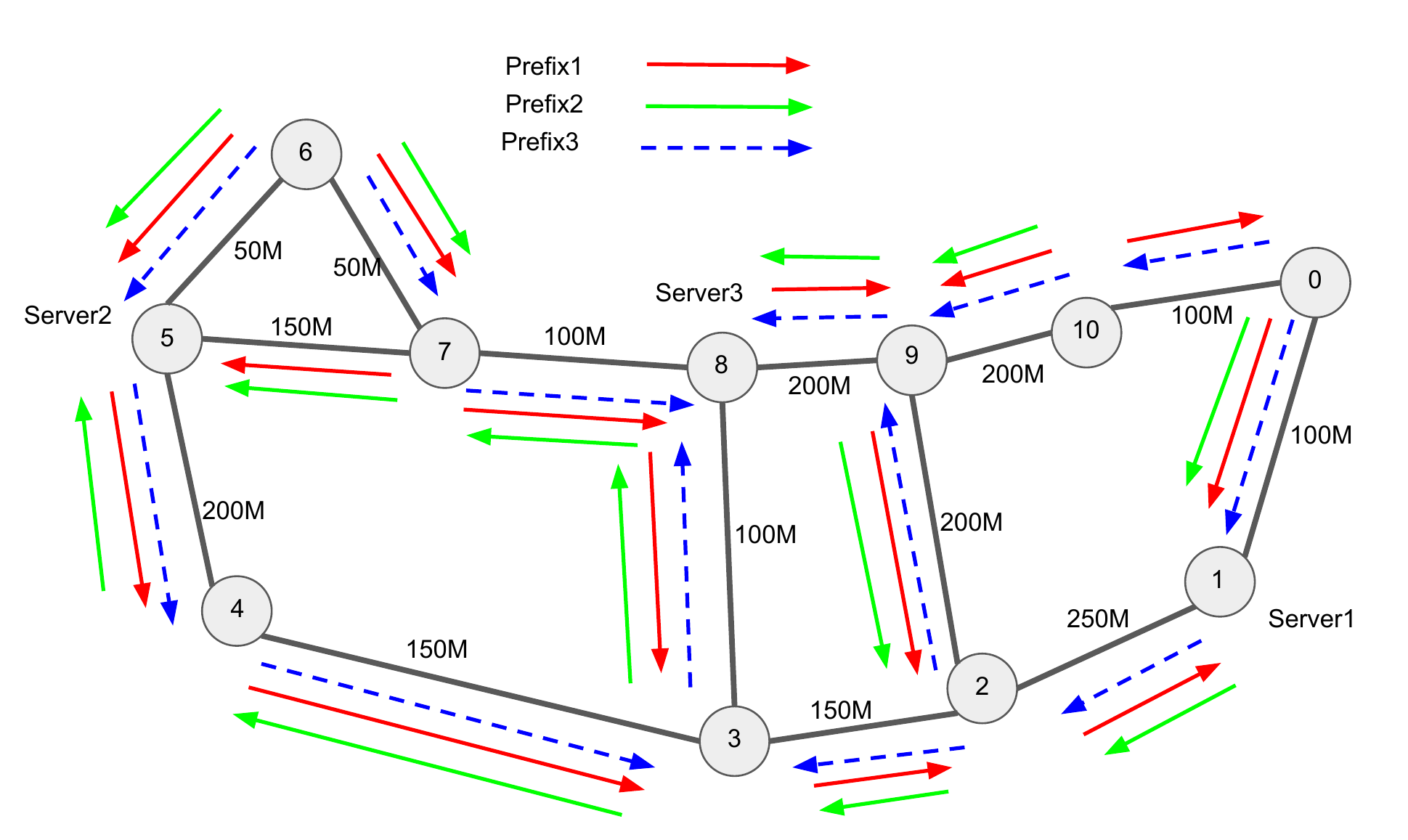}
			\caption{Abilene Topology \cite{giovanna}.}
			\label{topo:abilene}
		\end{subfigure}%
		\begin{subfigure}[t]{0.3\textwidth}
			\centering
			\includegraphics{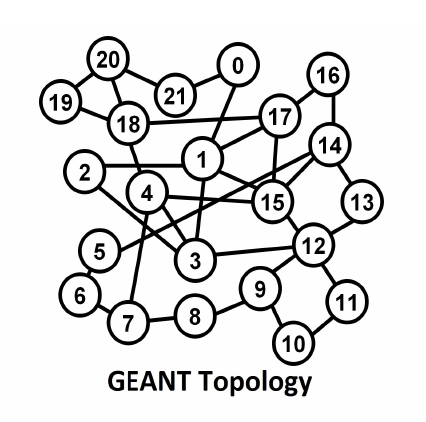}
			\caption{GEANT topology \cite{vip}.}
			\label{topo:geant}
		\end{subfigure}
		\begin{subfigure}[t]{0.3\textwidth}
			\centering
			\includegraphics{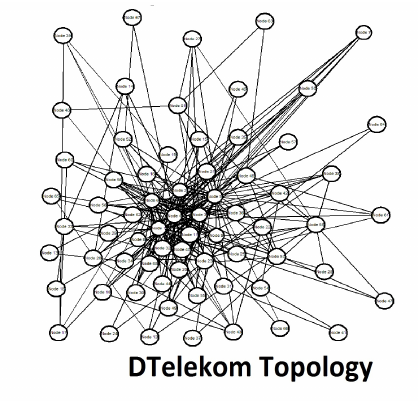}
			\caption{DTelekom topology \cite{vip}.}
			\label{topo:dt}
		\end{subfigure}	
		\begin{subfigure}[t]{0.3\textwidth}
			\centering
			\includegraphics[scale=0.45]{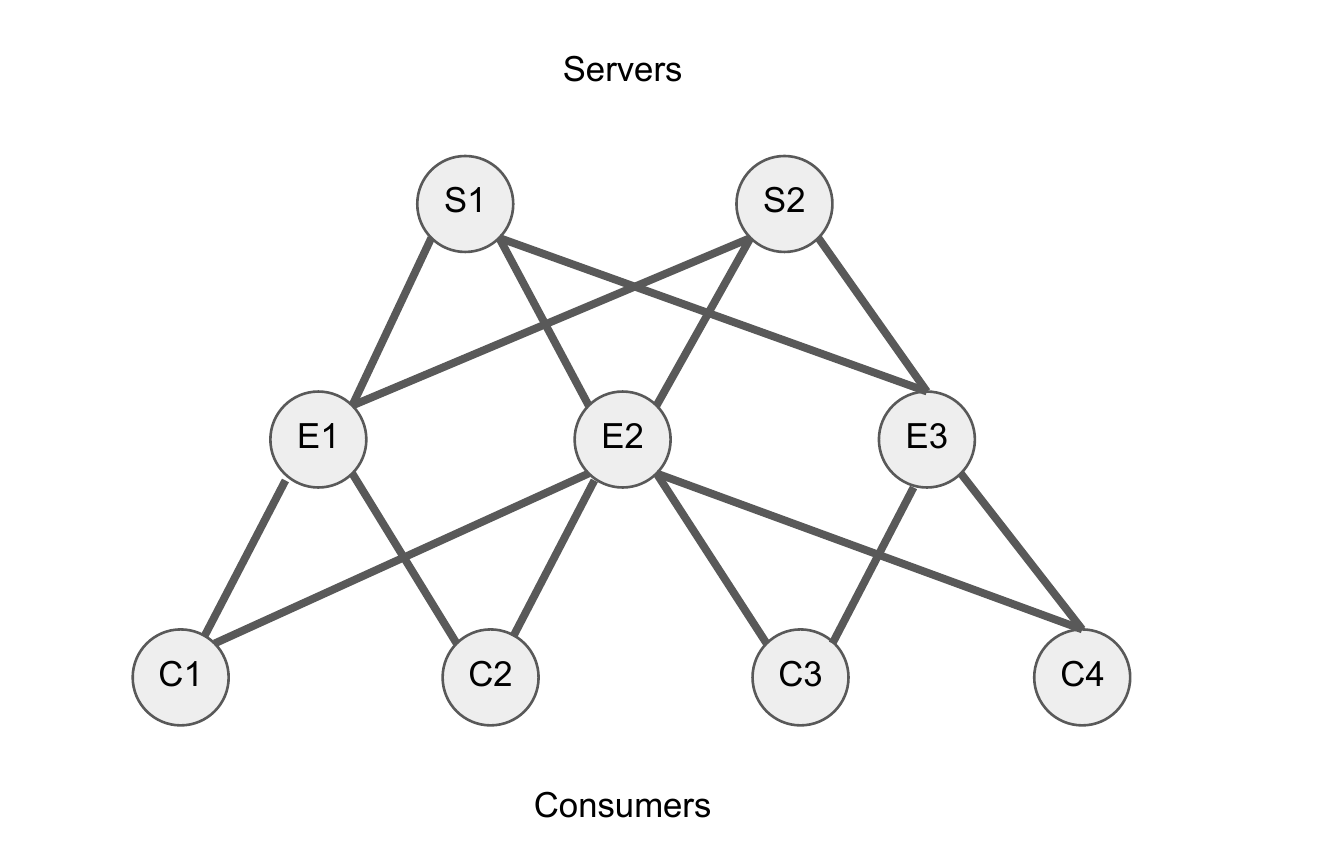}
			\caption{Tree topology.}
			\label{topo:tree}
		\end{subfigure}
		\begin{subfigure}[t]{0.3\textwidth}
			\centering
			\includegraphics[scale=.45]{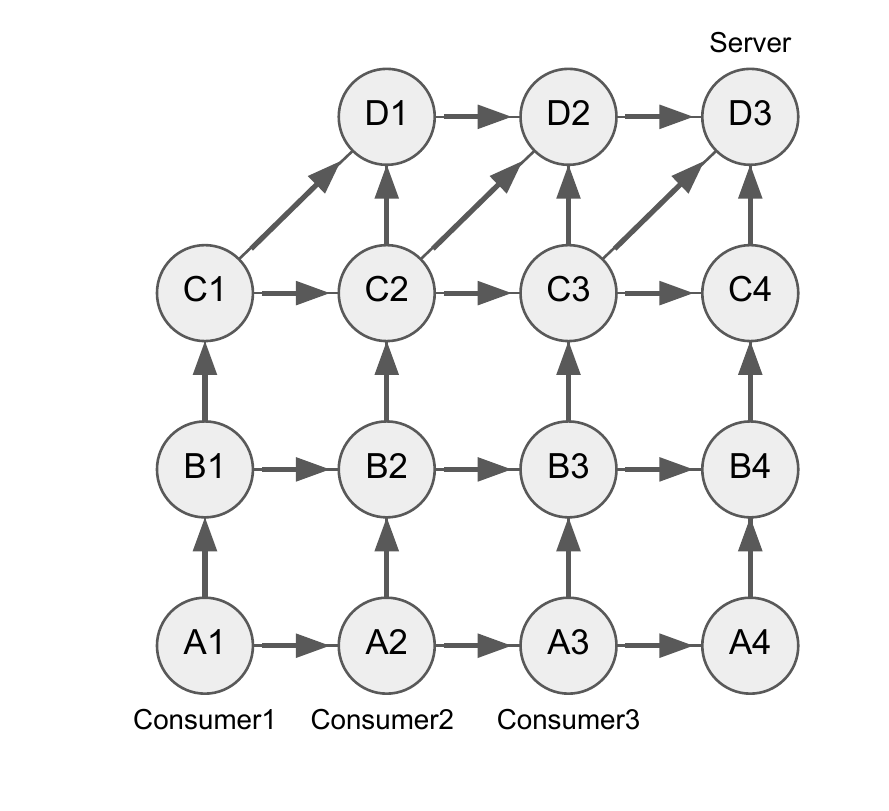}
			\caption{Ladder topology.}
			\label{topo:ladder}
		\end{subfigure}
		\begin{subfigure}[t]{0.3\textwidth}
			\centering
			\includegraphics[scale=.3]{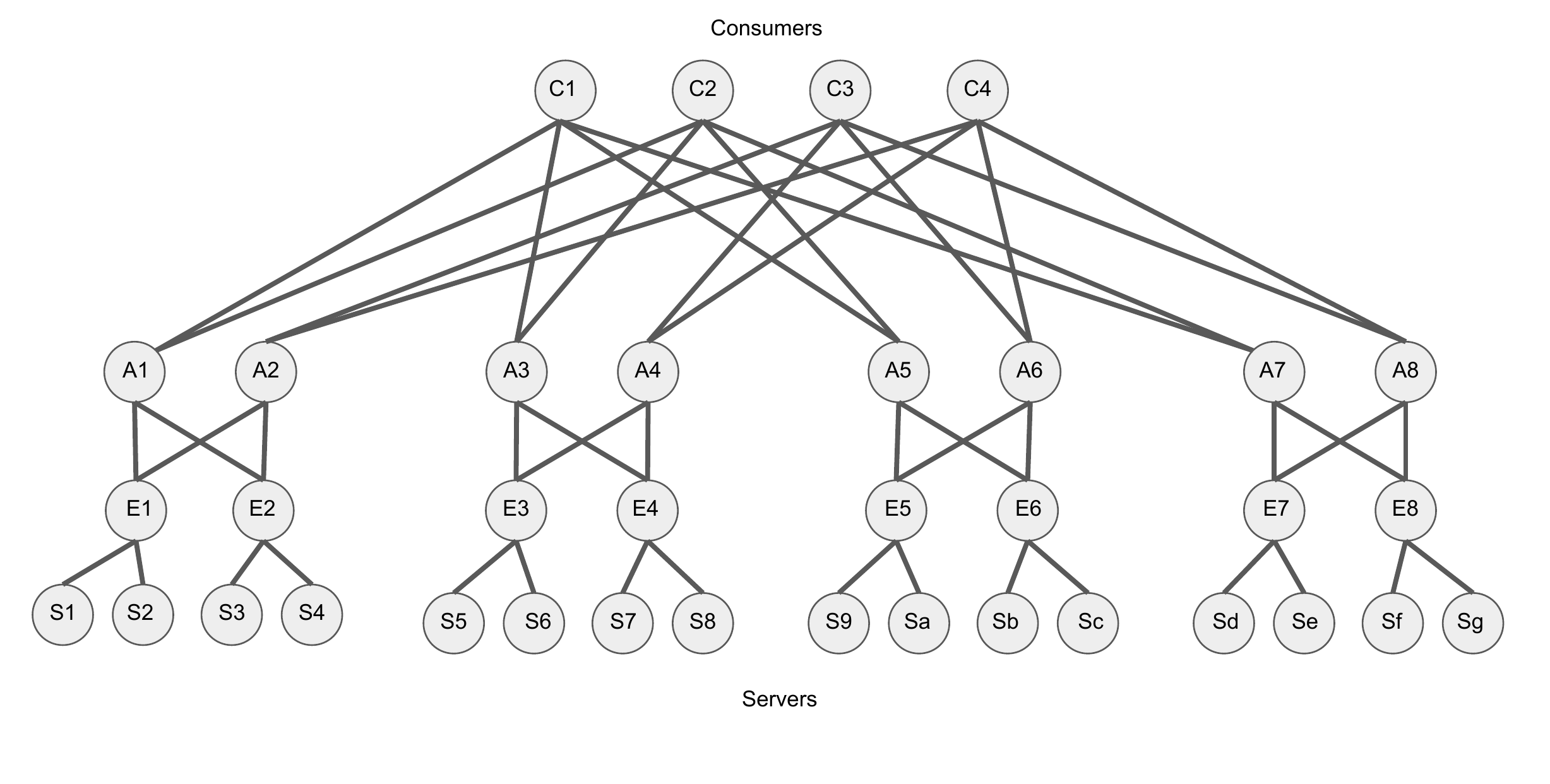}
			\caption{Fat Tree topology.}
			\label{topo:fattree}
		\end{subfigure}
		
		\caption{Network topologies used for the simulations.}\label{fig:topos}
	\end{figure*}		
	
	
	As mentioned above, the solutions $\bar{\rho}_i^n$ to (\ref{opt_rho}) are integer-avelued at each iteration.  However, for a general stepsize $a^n \in (0,1]$, the (relaxed) caching variables corresponding to the update in (\ref{conditional}) may not be integer-valued at each iteration. In particular, this would be true if the stepsize follows a diminishing stepsize rule.  Although one can explore rounding techniques and probabilistic caching techniques to obtain feasible integer-valued caching variables $x^n_i(k)$ from continuous-valued relaxed caching variables $\rho^n_i(k)$~\cite{stratis}, this would entail additional computational and communication complexity. 
	
	Since we are focused on distributed, low-complexity forwarding and caching algorithms, we require $\rho^n_i(k)$ to be either 0 or 1 at each iteration $n$.  This is realized by choosing the stepsize $a^n =1$ for all $n$. In this case, the update equation (\ref{conditional}) is reduced to:
	
	
	
	\begin{equation}
	\Phi^{n+1} = \bar{\Phi}^{n}. \label{conditional2}
	\end{equation}
	where $\bar{\Phi}^{n}$ is the solution to \eqref{opt_phi} and \eqref{opt_rho}.  That is, the solutions to the direction finding subproblems provide us with forwarding and caching decisions at each iteration.  We now summarize the remarkably elegant MinDelay forwarding and caching algorithm.
	
	{\bf MinDelay Forwarding Algorithm}: At each iteration $n$, each node $i$ and for each object $k$, the forwarding algorithm chooses the outgoing link $(i,m)$ to forward requests for object $k$, where $m$ is chosen according to
	\begin{equation} \label{forwarding}
	m \in \arg \min_{j\in\mathcal{O}(i,k)} \delta^n_{ij}(k). 
	\end{equation}
	That is, requests for object $k$ are forwarded on an outgoing link with the minimum marginal forwarding cost. 
	
	{\bf MinDelay Caching Algorithm}:  At each iteration $n$, each node $i$ calculates a cache score $CS^n(i,k)$ for each object $k$ according to
	\begin{equation}
	CS^n(i,k) \triangleq t^n_i(k)\delta^n_{i}(k).\label{cache_score}
	\end{equation}
	where $\delta^n_{i}(k) \equiv \min_{j\in\mathcal{O}(i,k)} \delta^n_{ij}(k)$.  Upon reception of data object $k_{new}$ not currently in the cache of node $i$, if the cache is not full, then $k_{new}$ is cached.  If the cache is full, then $CS^n(i,k_{new})$ is computed, and compared to the lowest cache score among the currently cached objects, denoted by $CS^n(i,k_{min})$.  If $CS^n(i,k_{new}) > CS^n(i,k_{min})$, then replace $k_{min}$ with $k_{new}$.  Otherwise, the cache contents stay the same. 
	
	The cache score given in (\ref{cache_score}) for a given content $k$ at node $i$ is the minimum marginal forwarding cost for object $k$ at $i$, multiplied by the total request rate for $k$ at $i$. By caching the data objects with the highest cache scores, each node maximally reduces the total cost of forwarding request traffic.

	
	
	One drawback of using stepsize $a^n=1$ in the MinDelay algorithm is that it makes studying the asymptotic behavior of the algorithm difficult.  Nevertheless, in extensive simulations shown in the next section, we observe that the algorithm behaves in a stable manner asymptotically.  Moreover, the MinDelay significantly outperform several state-of-the-art caching and forwarding algorithms in important operating regimes.  

	\begin{figure*}
		\begin{subfigure}[t]{0.5\textwidth}
			\centering
			\includegraphics[width=1\textwidth,height=.65\textwidth]{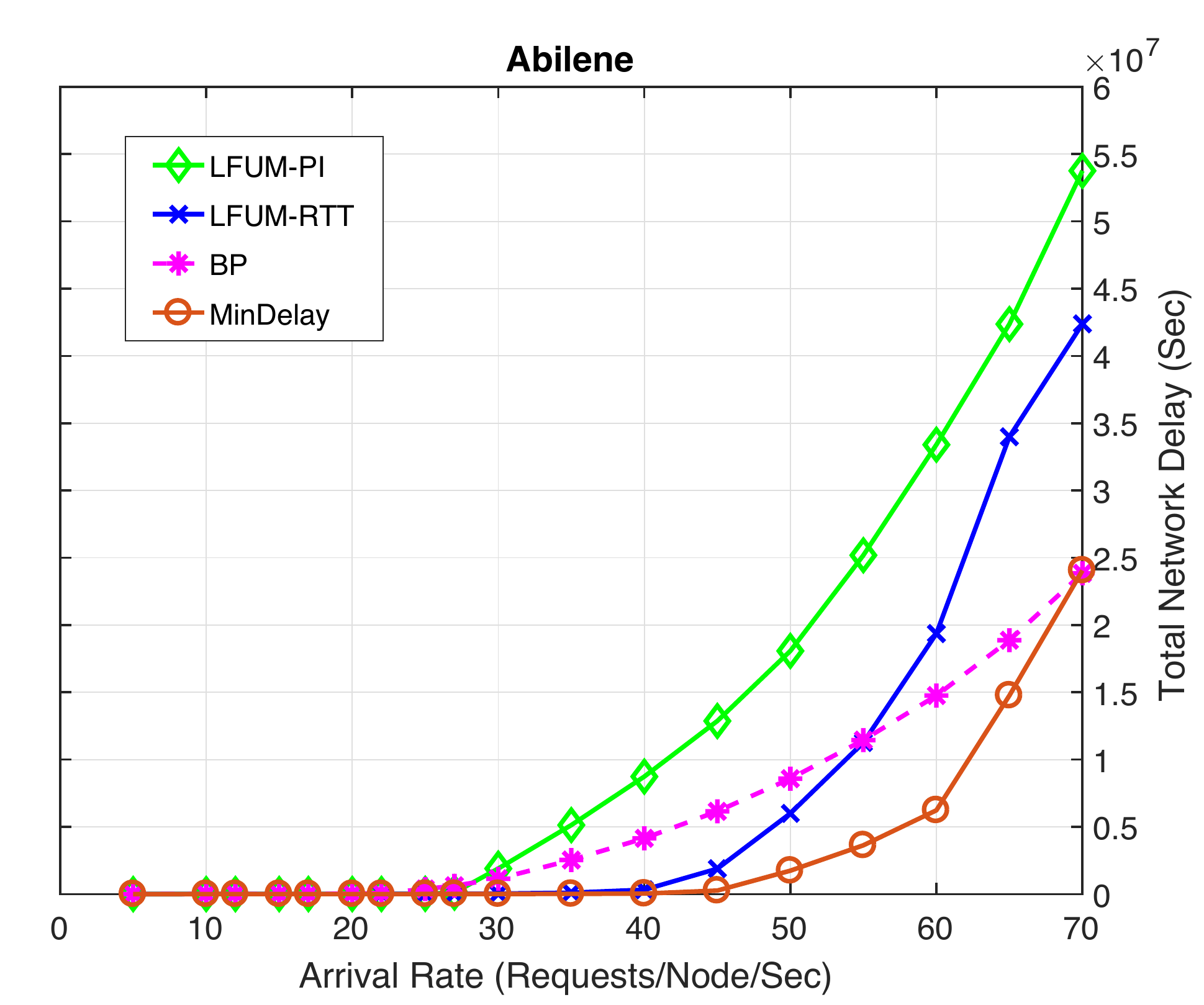}
			\caption{Abilene Topology.}
			\label{fig:abilene}
		\end{subfigure}%
		\begin{subfigure}[t]{0.5\textwidth}
			\centering
			\includegraphics[width=1\textwidth,height=.65\textwidth]{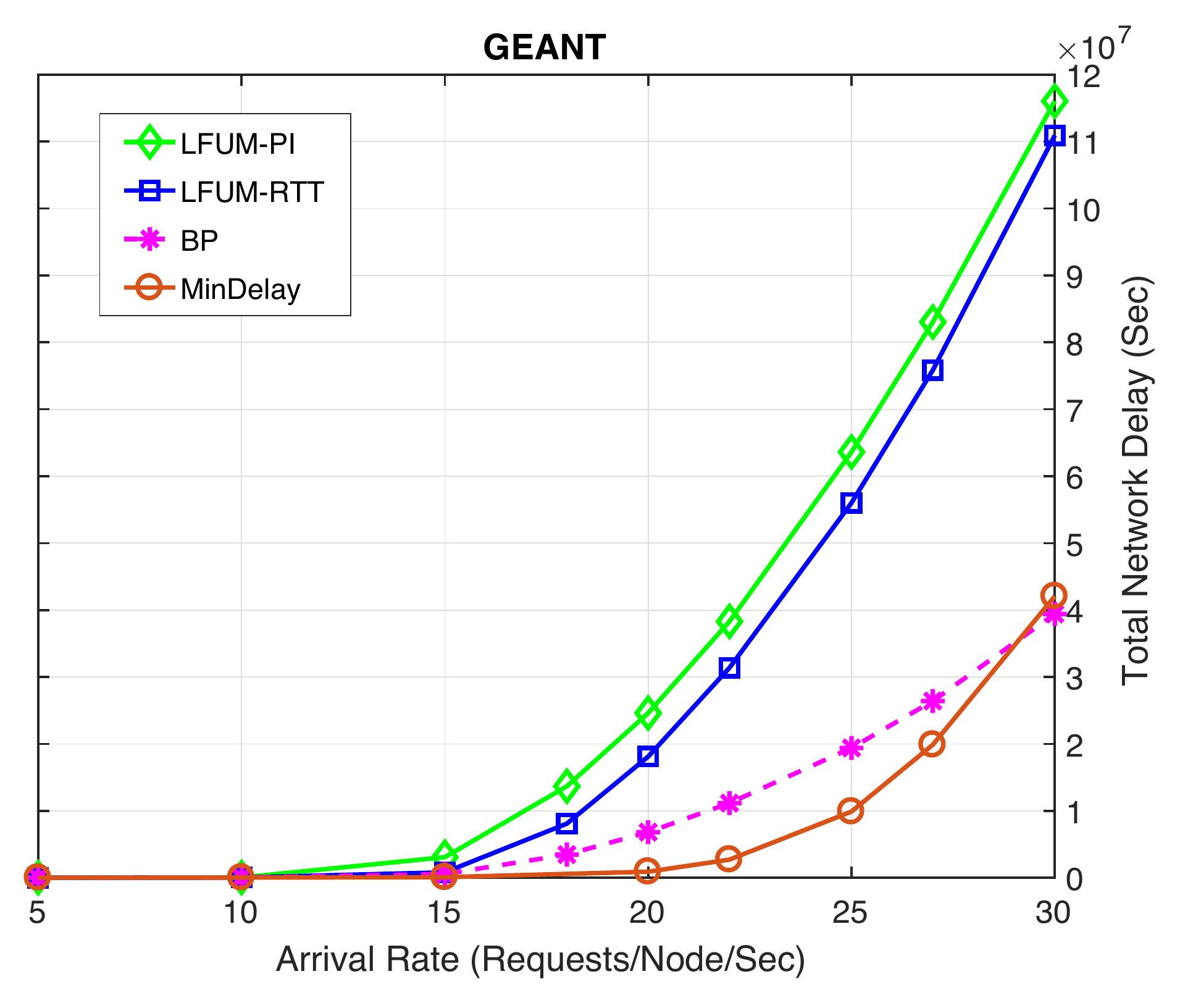}
			\caption{GEANT topology.}
			\label{fig:geant}
		\end{subfigure}
		\begin{subfigure}[t]{0.5\textwidth}
			\centering
			\includegraphics[width=1\textwidth,height=.65\textwidth]{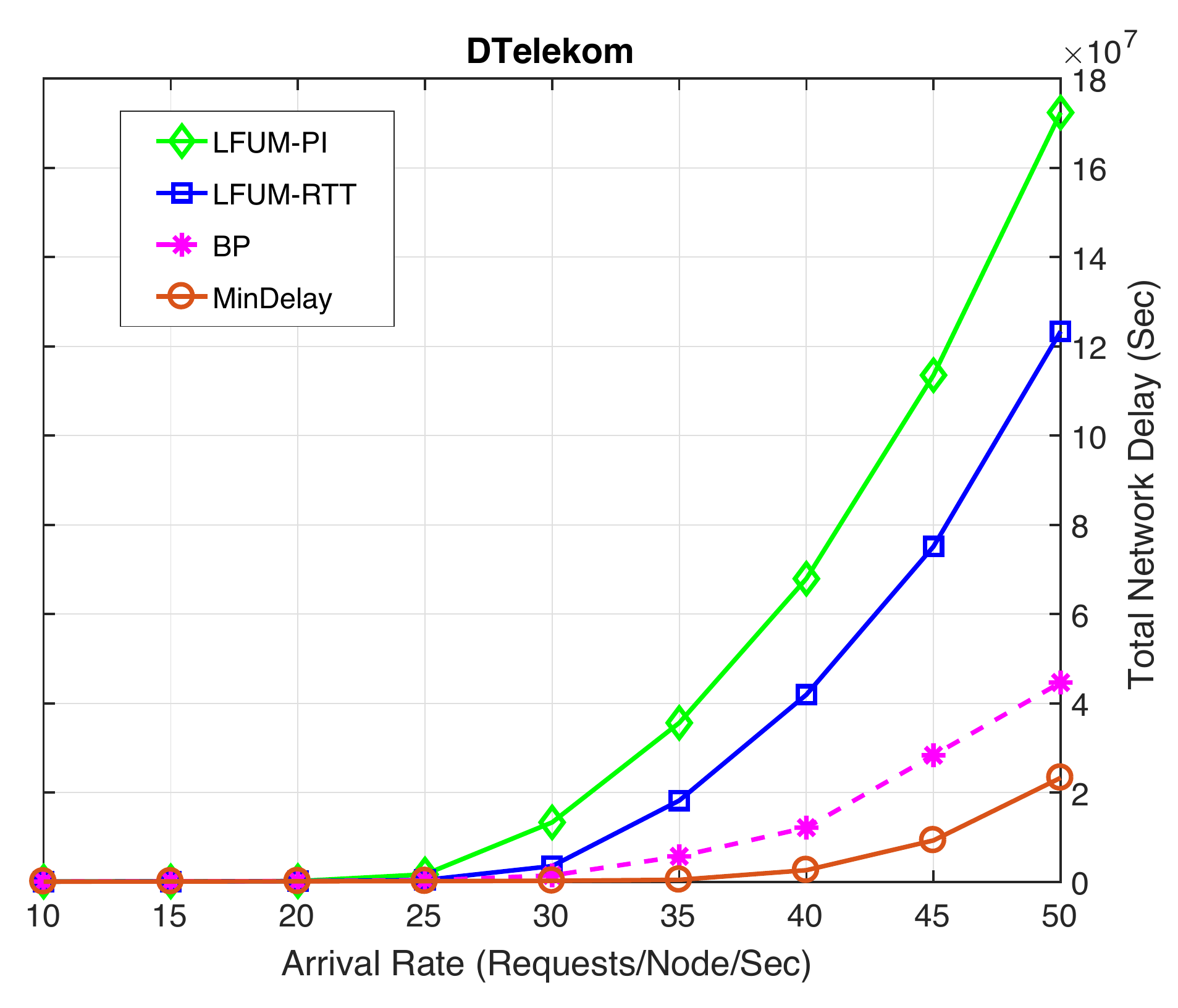}
			\caption{DTelekom topology.}
			\label{fig:dt}
		\end{subfigure}	
		\begin{subfigure}[t]{0.5\textwidth}
			\centering
			\includegraphics[width=1\textwidth,height=.65\textwidth]{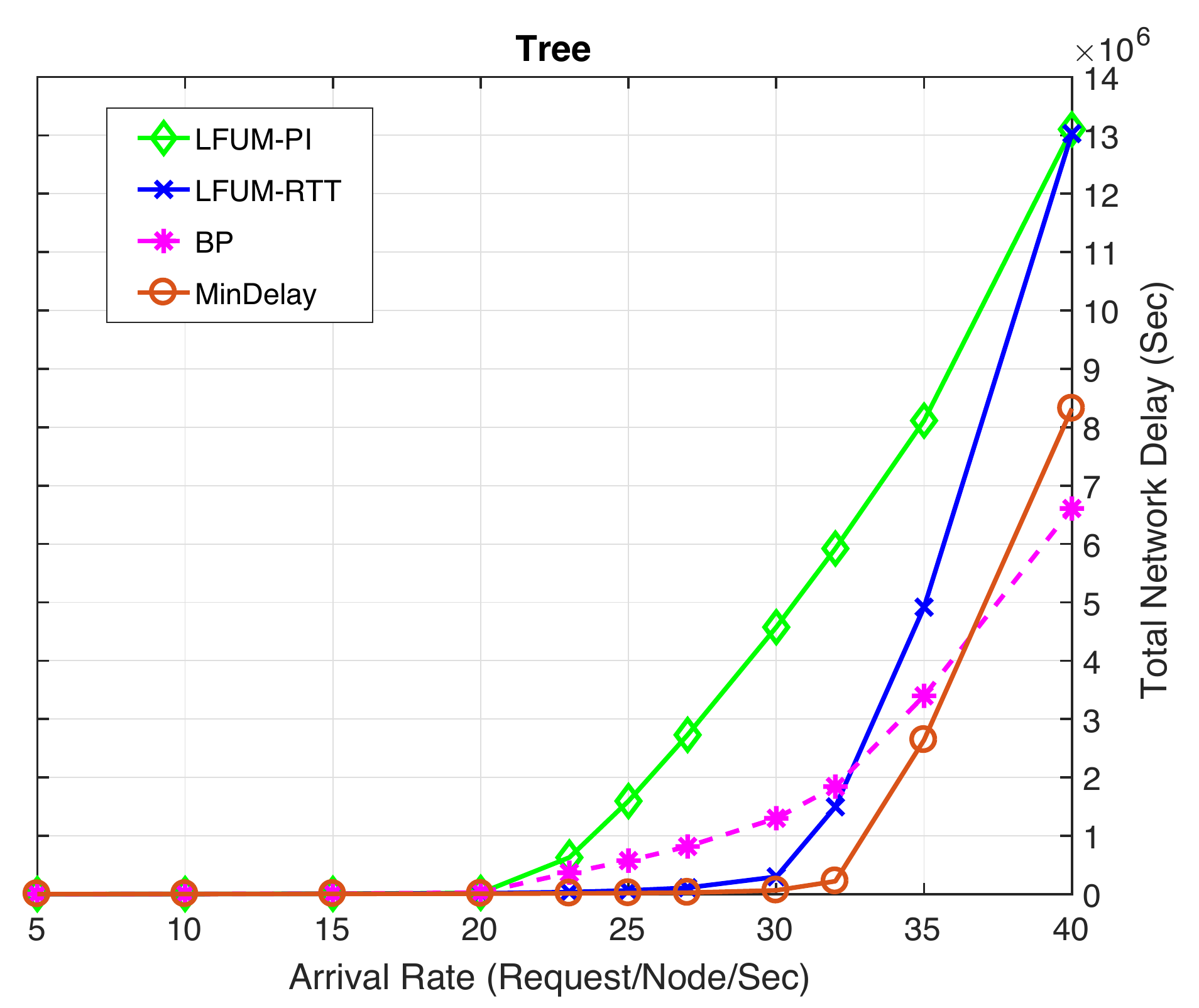}
			\caption{Tree topology.}
			\label{fig:tree}
		\end{subfigure}
		\begin{subfigure}[t]{0.5\textwidth}
			\centering
			\includegraphics[width=1\textwidth,height=.65\textwidth]{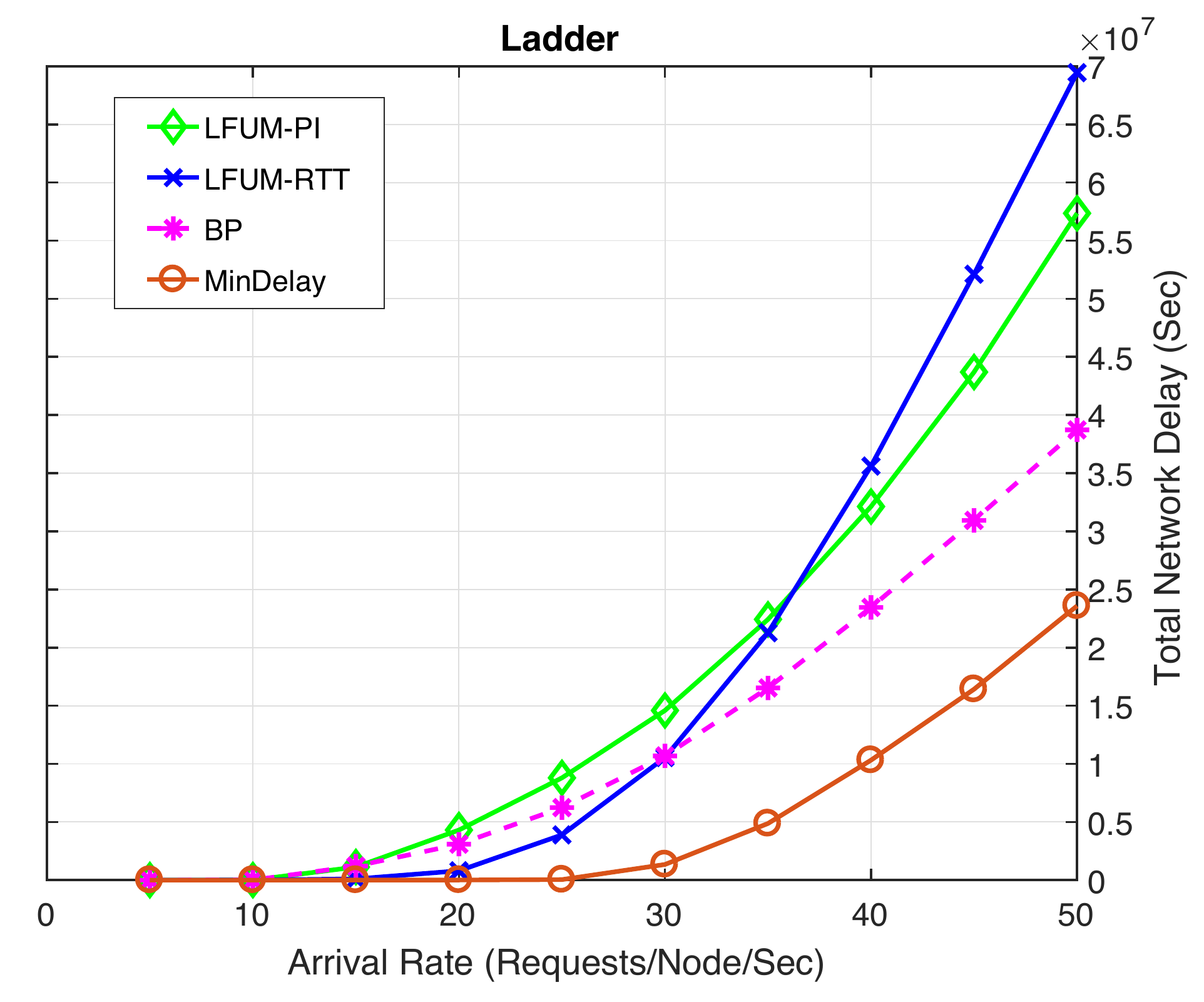}
			\caption{Ladder topology.}
			\label{fig:ladder}
		\end{subfigure}
		\begin{subfigure}[t]{0.5\textwidth}
			\centering
			\includegraphics[width=1\textwidth,height=.65\textwidth]{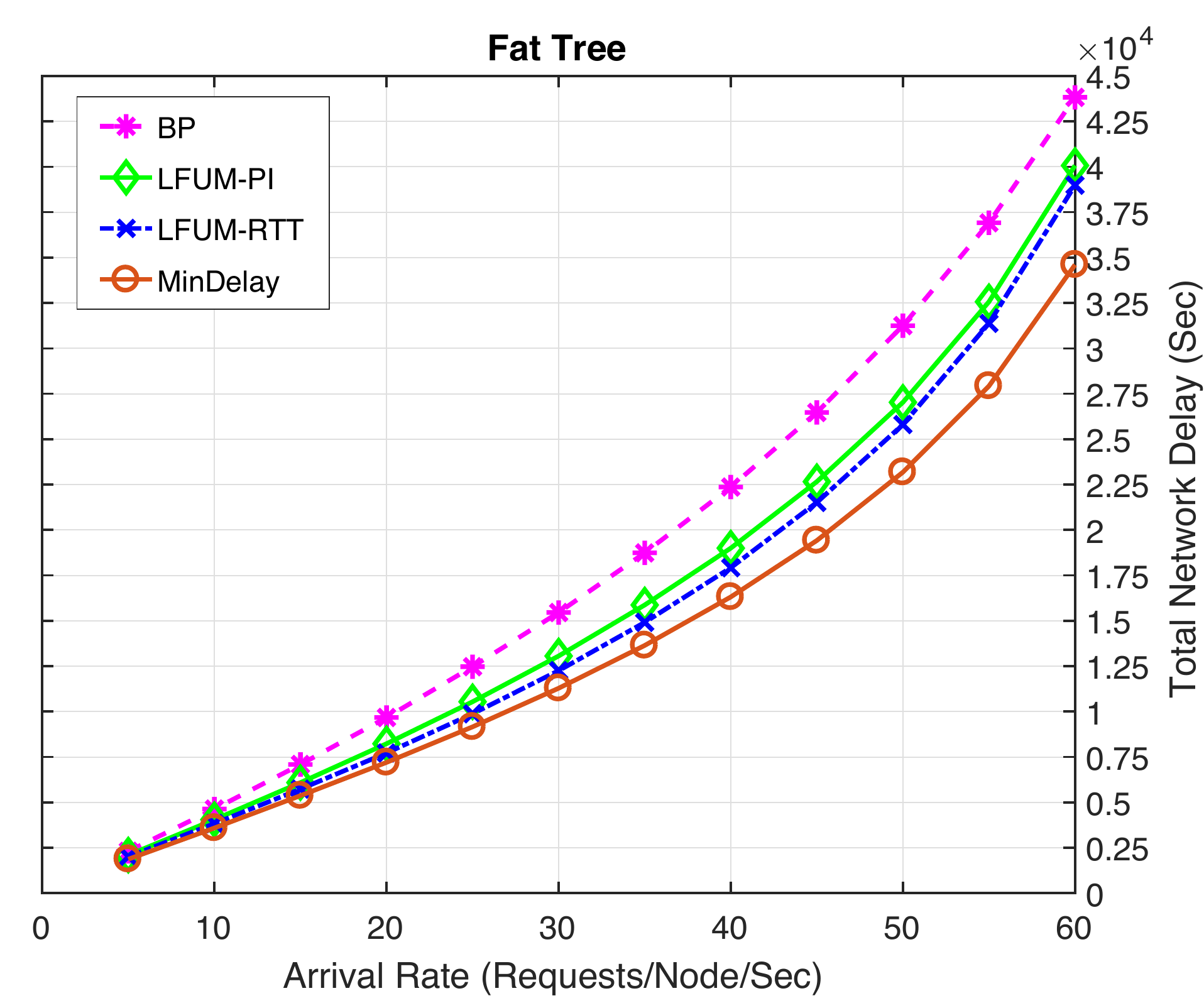}
			\caption{Fat Tree topology.}
			\label{fig:fattree}
		\end{subfigure}
		
		\caption{Total network delay (sec) vs. request arrival rate (requests/node/sec).}\label{fig:results}
	\end{figure*}	
	\begin{figure*}
		\begin{subfigure}[t]{0.48\textwidth}
			\centering
			\includegraphics[width=1\textwidth,height=0.65\textwidth]{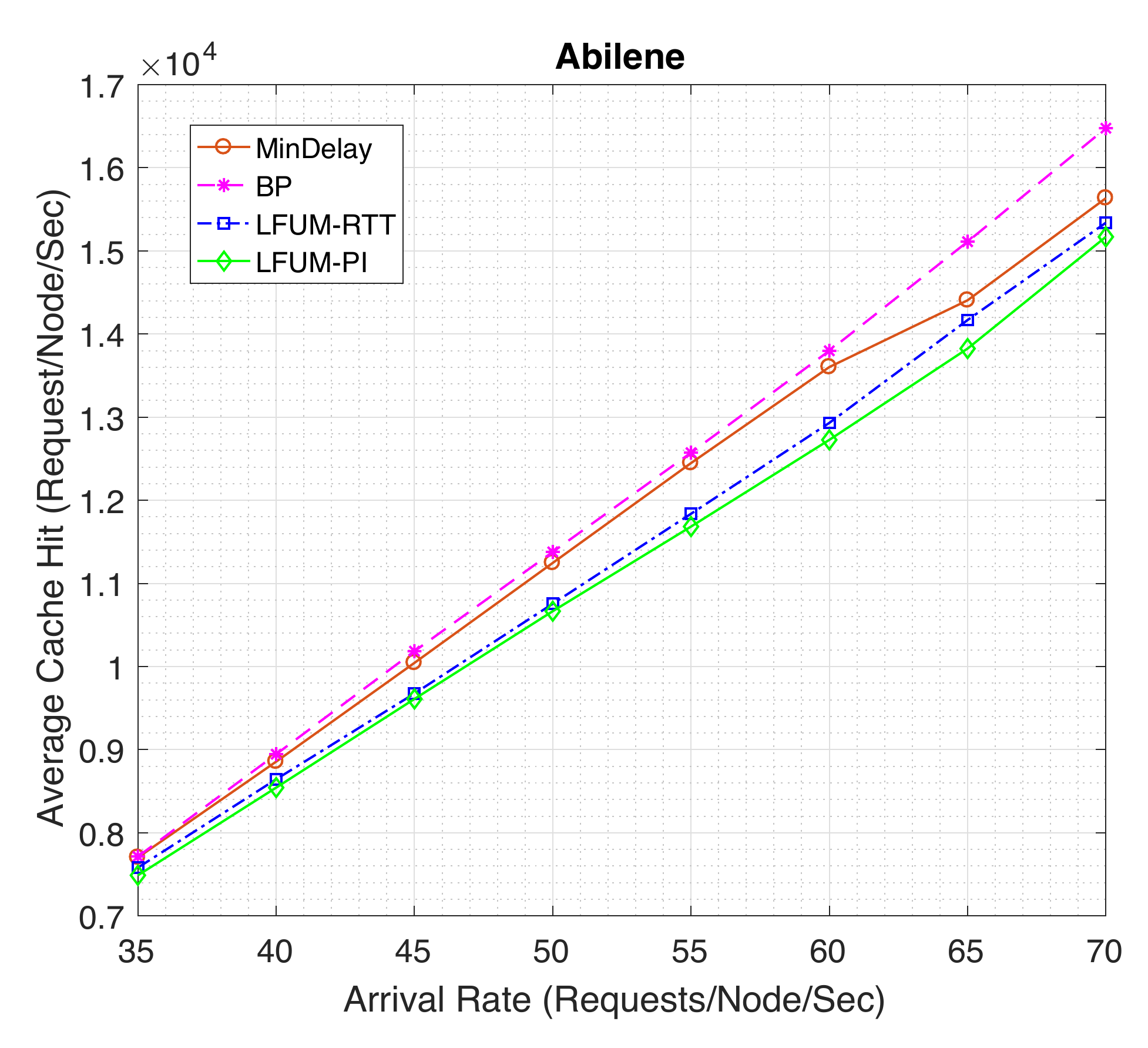}
			\caption{Abilene Topology.}
			\label{fig:abilene-ch}
		\end{subfigure}%
		\begin{subfigure}[t]{0.48\textwidth}
			\centering
			\includegraphics[width=1\textwidth,height=0.65\textwidth]{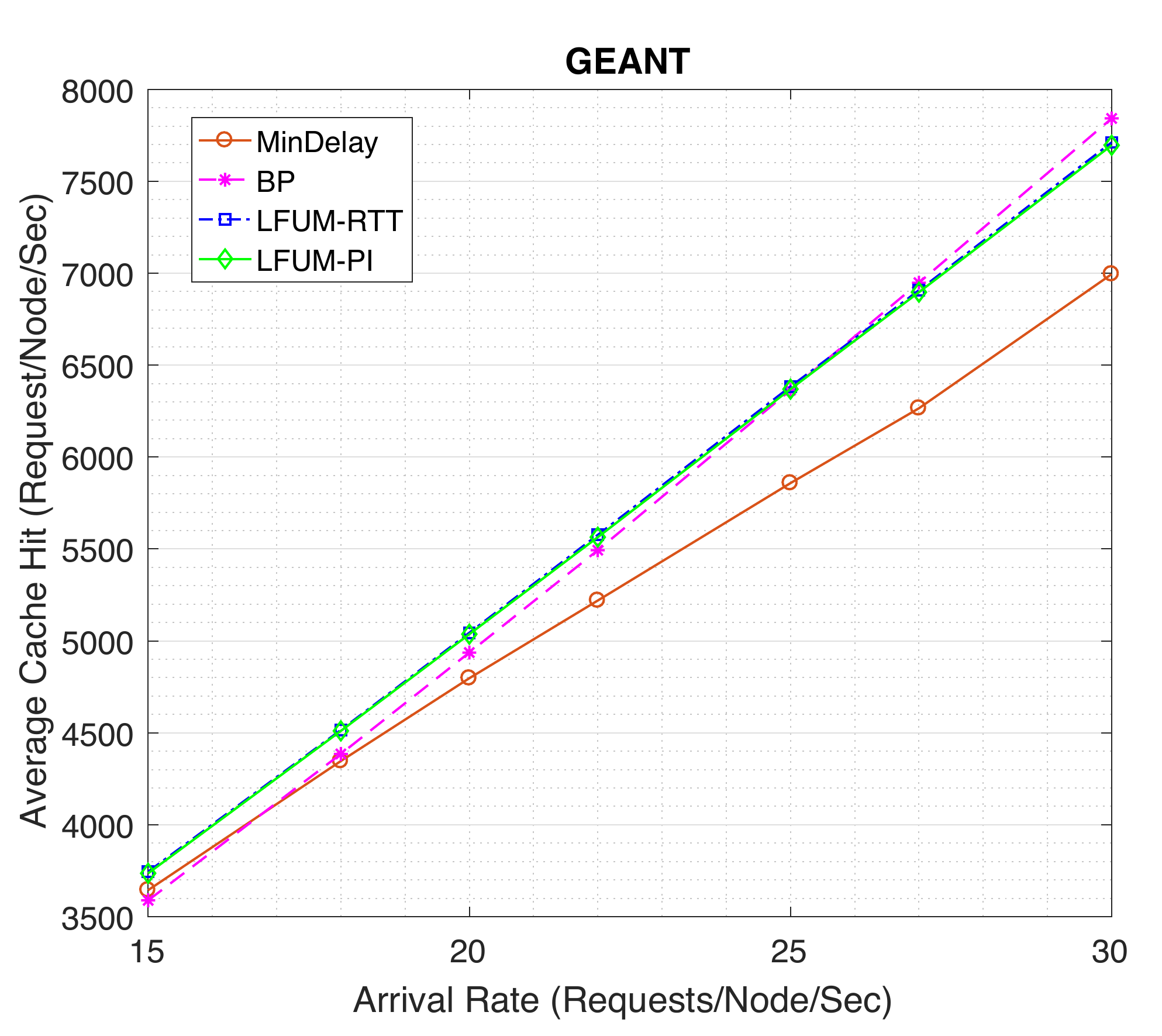}
			\caption{GEANT topology.}
			\label{fig:geant-ch}
		\end{subfigure}
		
		\begin{subfigure}[t]{0.48\textwidth}
			\centering
			\includegraphics[width=1\textwidth,height=0.65\textwidth]{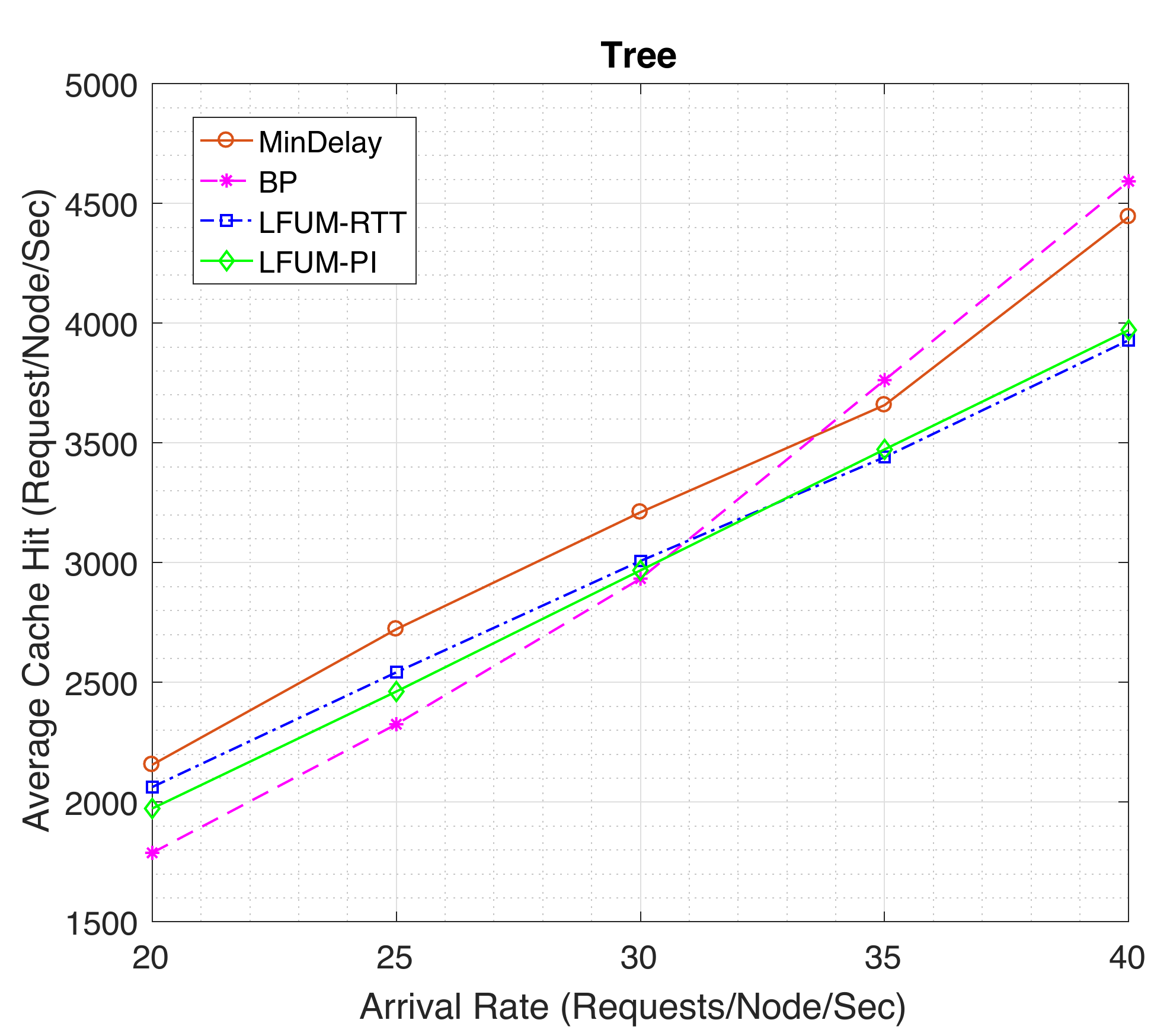}
			\caption{Tree topology.}
			\label{fig:tree-ch}
		\end{subfigure}
		\begin{subfigure}[t]{0.48\textwidth}
			\centering
			\includegraphics[width=1\textwidth,height=0.65\textwidth]{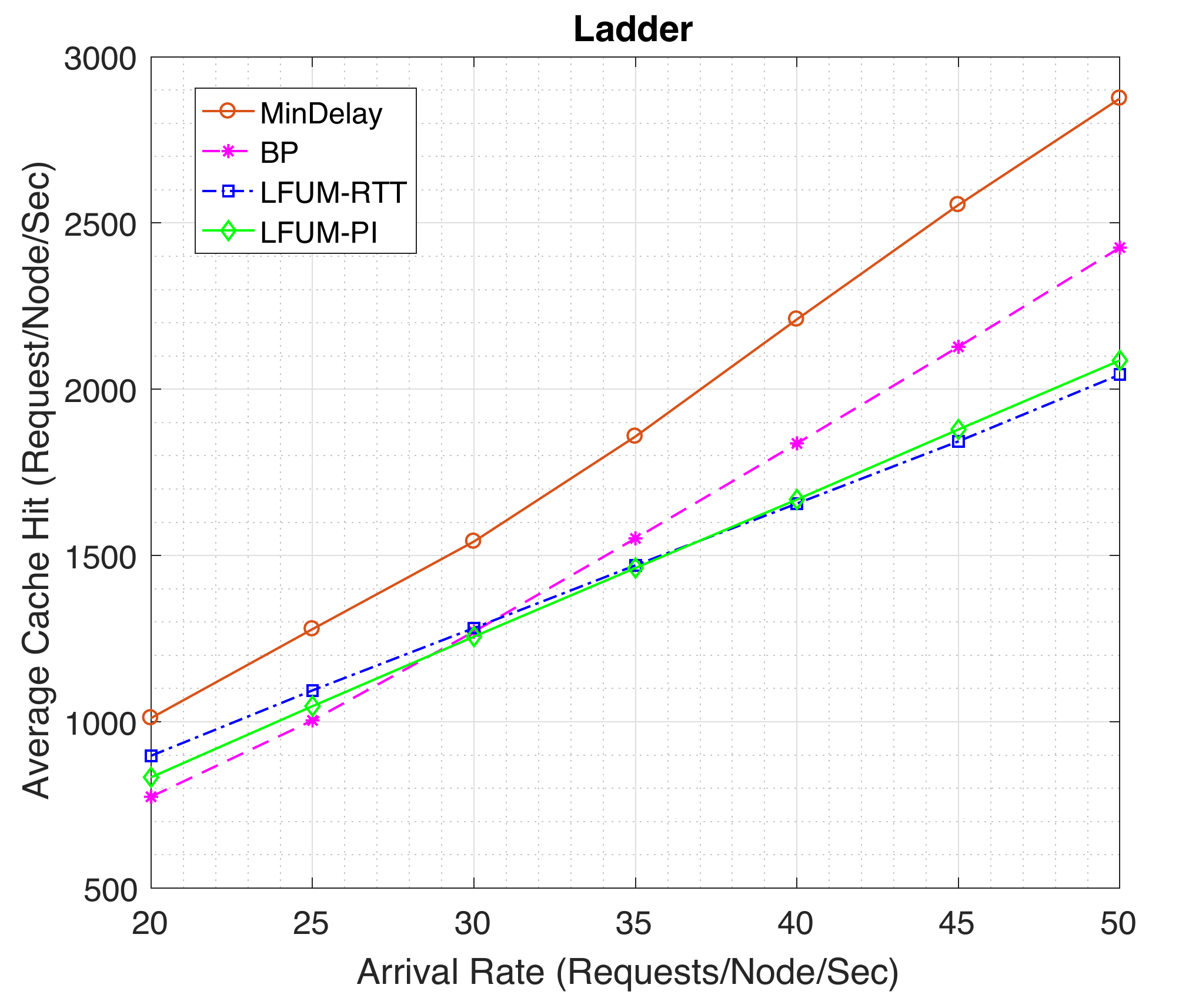}
			\caption{Ladder topology.}
			\label{fig:ladder-ch}
		\end{subfigure}
		
		\caption{Average total cache hits (requests/node/sec) vs. Arrival rate (requests/node/sec).}\label{fig:results-ch}
	\end{figure*}

	\section{Simulation Experiments}
	\label{sec:simulation}
	
	In this section we present the results of extensive simulations performed using our own Java-based ICN Simulator.  We have considered three competing schemes for comparison with MinDelay.  First, we consider the VIP joint caching and forwarding algorithm introduced in~\cite{vip}.  This algorithm uses a backpressure(BP)-based scheme for forwarding and a stable caching algorithm, both based on VIP (Virtual Interest Packet) queue states~\cite{vip}.  In the VIP algorithm discussed in~\cite{vip}, multiple Interest Packets requesting the same Data Packet are aggregated.  Since we do not consider Interest Packet aggregation in this paper, we compare MinDelay with a version of VIP without Interest aggregation, labeled BP.  We consider the VIP algorithm (or BP) to be the direct competitor with MinDelay, since to the best of our knowledge, it is the only other scheme that explicitly jointly optimizes forwarding and caching for general ICN networks.  
	
	The other two approaches implemented here are based on the LFU cache eviction policy. We note that for stationary input request processes, the performance of LFU is typically much better than those of LRU and FIFO.\footnote{Initially we included LRU-based approaches.  However, since their performance was much worse than the competitors, we omitted them in the final figures.} In the first approach, denoted by LFUM-PI, multipath request forwarding is based on the scheme proposed in \cite{giovanna}.  Here, the forwarding decision is made as follows: an Interest Packet requesting a given object is forwarded on an outgoing interface with a probability inversely proportional to the number of Pending Interest (PI) Packets for that object on that outgoing interface.  The second LFU-based approach implemented here, denoted by LFUM-RTT, has a RTT-based forwarding strategy. In this strategy, described in \cite{Detti-modeling}, the multipath forwarding decision is based on the exponentially weighted moving average of the RTT of each outgoing interface per object name. An Interest Packet requesting an object is forwarded on an outgoing interface with a probability inversely proportional to the average RTT recorded for that object on that outgoing interface.
	
	We tested the MinDelay forwarding and caching algorithm against the described approaches on several well-known topologies depicted in Fig. \ref{fig:topos}. In the following, we explain the simulation scenarios and results in detail. 
	
	\subsection{Simulation Details}
	
	Each simulation generates requests for 1000 seconds and terminates when all the requested packets are fulfilled. During the simulation, a requesting node requests a content object by generating an Interest Packet containing the content name and a random nonce value, and then submits it to the local forwarder. Upon reception of an Interest Packet, the forwarder first checks if the requested content name contained in the Interest Packet is cached in its local storage. If there is a copy of the content object in the local storage, it generates a Data Packet containing the requested object, along with the content name and the nonce value, and puts the Data Packet in the queue of the interface on which the Interest Packet was received. If the local cache does not have a copy of the requested object, the forwarder uses the FIB table to retrieve the available outgoing interfaces.\footnote{In the simulations, we ensured that loop-free routing was done prior to the forwarding and caching experiments.  The results of the routing algorithm are saved in FIB tables at each node.} Then, the forwarder selects an interface among the available interfaces based on the implemented forwarding strategy. In particular, for MinDelay, we update the marginal forwarding costs given in \eqref{deltanijk} at the beginning of each update interval (with a length between 2-5 seconds), and cache the results in a sorted array for future use. Hence, the forwarding decision given in \eqref{forwarding} takes O(1) operations. 
	
	After selecting the interface based on the considered forwarding strategy, the forwarder creates a Pending Interest Table (PIT) entry with the key being the content name concatenated with the nonce value, and the PIT entry value being the incoming interface ID. Note that we concatenate the nonce value to the content name since we do not assume Interest Packet suppression at the forwarder.  Hence, we need to have distinguishable keys for each Interest Packet.  Next, the forwarder assigns the Interest Packet to the queue of the selected interface, to be transmitted in a FIFO manner.  
	
	Upon reception of a Data Packet, the forwarder first checks if the local storage is full. If the storage is not full, it will cache the contained data object~\footnote{In the experiments, all data objects contain one chunk, or one Data Packet.} in local storage. If the storage is at capacity, it uses the considered cache eviction policy to decide whether to evict an old object and replace it with the new one.  In the case of MinDelay, the forwarder regularly updates the cache score of the currently-cached contents using \eqref{cache_score} at the beginning of the update intervals and keeps a sorted list of the cached content objects using a hash table and a priority queue. When a new Data Packet arrives, the forwarder computes its cache score, and compares the score with the lowest cache score among the currently-cached content objects. If the score of the incoming Data Packet is higher than the current lowest cache score, the forwarder replaces the corresponding cached object with the incoming one.  Otherwise, the cached contents remain the same.
	
	Finally, the forwarder proceeds by retrieving and removing the PIT entry corresponding to the Data Packet and assigning the Data Packet to the queue of the interface recorded in the PIT entry.
	
	In all topologies, the number of content objects is 5000. Each requester requests a content object according to a Zipf distribution with power exponent $\alpha=0.75$, by generating an Interest Packet each of size 1.25 KBytes. All content objects are assumed to have the same size and can be packaged into a single Data Packet of size 500 KBytes. The link capacity of all the topologies, except in Abilene topology illustrated in Fig. \ref{topo:abilene}, is 50 Mbps. 
	
	We first consider the Abilene topology \cite{giovanna} depicted in Figure \ref{topo:abilene}.  There are three servers, at nodes 1, 5, and 8, each serving 1/3 of the content objects. That is, object $k$ is served by server $k\text{ mod } 3 +1$ for $k=1,2,\ldots,5000$.  The other eight nodes of the topology request objects according to Zipf distribution with $\alpha =0.75$. Also, each requester has a content store of size 250 MBytes, or equivalently 500 content objects. 
	
	In the GEANT topology, illustrated in Figure \ref{topo:geant}, there are 22 nodes in the network. All nodes request content objects. Each content object is randomly assigned to one of the 22 nodes as its source node. Each node has a content store of size 250 MBytes, or equivalently 500 content objects. 
	
	In the DTelekom topology , illustrated in Figure \ref{topo:dt}, there are 68 nodes in the network.  All nodes request content objects. Each content object is randomly assigned to one of the 68 nodes as its source node. Each node has a content store of size 250 MBytes, or equivalently 500 content objects. 
	
	In the Tree topology, depicted in Figure \ref{topo:tree}, there are four requesting nodes at the leaves, C1, C2, C3 and C4. There are three edge nodes, E1, E2, and E3. Each content object is randomly assigned to one of the two source nodes, S1 and S2. Each requesting and edge node has a content store of size 125 MBytes, or equivalently 250 content objects. 
	
	In the Ladder topology \cite{giovanna}, depicted in Figure \ref{topo:ladder}, there are three requesting nodes, A1, A2 and A3. The source of all the content objects are at node D3. Each node in the network, except node D3, has a content store of size 125 MBytes, or equivalently 250 content objects. 
	
	Finally, in the Fat Tree topology, depicted in Figure \ref{topo:fattree}, requesters are at the roots, i.e., nodes C1, C2, C3 and C4. There are 16 servers at the leaves. In this topology, each content object is randomly assigned to two servers, one chosen from the first 8 servers, and the other from the second 8 servers. All the requesting nodes as well as Aggregation and Edge nodes have a content store, each of size 125 MBytes, or equivalently 250 content objects. 
	
	\subsection{Simulation Results}
	
	In Figures \ref{fig:results} and \ref{fig:results-ch}, the results of the simulations are plotted.  The figures illustrate the performance of the implemented schemes in terms of total network delay for satisfying all generated requests (in seconds) and the average cache hits in requests/node/second, versus the arrival rate in requests/node/second, respectively. We define the delay for a request as the difference between the creation time of the Interest Packet and the arrival time of its corresponding Data Packet at the requesting node. A cache hit for a data object is recorded when an Interest Packet reaches a node which is not a content source but which has the data object in its cache. When a cache hit occurs, the corresponding metric is incremented one. 
	
	To reduce randomness in our results, we ran each simulation 10 times, each with a different seed number, and plotted the average performance of each scheme in Figures \ref{fig:results} and \ref{fig:results-ch}. 
	
	Figure \ref{fig:results} shows the total network delay in seconds versus the per-node arrival rate in request/seconds, for the above-mentioned topologies.  As can be seen, in all the considered topologies, MinDelay has lower delay in the low to moderate arrival rate regions.  In the higher arrival rate regions, BP's outperforms MinDelay in 3 of the tested topologies (Abilene, GEANT, and Tree), 

	As shown in~\cite{vip}, the BP performs well in high arrival rate regions since the VIP algorithm adaptively maximizes the throughput of Interest Packets, thereby maximizing the stability region of user demand rates satisfied by the network.  When the network is operating well within the stability region, however, MinDelay typically has superior performance.  Thus, the MinDelay and VIP algorithms complement each other in delivering superior delay performance across the entire range of request arrival rates.


	Finally, Figure \ref{fig:results-ch} depicts the average total cache hits of the network (in requests/node/second) versus the per-node arrival rate (in request/seconds) for the Abilene, GEANT, Tree, and Ladder topologies, respectively. It can be seen that the cache hit performance of MinDelay is competitive but not necessarily superior to those of the other algorithms.  This follows form the fact that MinDelay is designed with the objective of decreasing total network delay, and not explicitly with the objective of increasing cache hits.
	
	\section{Conclusion}
	
	In this work, we established a new  unified framework for minimizing congestion-dependent network cost by jointly choosing node-based forwarding and caching variables.  Relaxing integer constraints on caching variables, we used a version of the conditional gradient algorithm to develop MinDelay,  an adaptive and distributed joint forwarding and caching 
	algorithm for the original mixed-integer optimization problem.  The MinDelay algorithm elegantly yields feasible routing variables and integer caching variables at each iteration, and can be implemented in a distributed manner with low complexity and overhead. 
	
	Simulation results show that while the VIP algorithm performs well in high  request arrival rate regions, MinDelay typically has significantly better delay performance in the low to moderate request rate regions.  Thus, the MinDelay and VIP algorithms complement each other in delivering superior delay performance across the entire range of request arrival rates.
	
	The elegant simplicity and superior performance of the MinDelay algorithm raise many interesting questions for future work.  Specifically, we are interested in analytically characterizing the time-asymptotic behavior of MinDelay, as well as providing guarantees on the gap between the MinDelay performance and the theoretically optimal performance for the joint forwarding and caching problem.

	\bibliographystyle{IEEEtran}
	\bibliography{OptRef}

\begin{thebibliography}{10}
\providecommand{\url}[1]{#1}
\csname url@samestyle\endcsname
\providecommand{\newblock}{\relax}
\providecommand{\bibinfo}[2]{#2}
\providecommand{\BIBentrySTDinterwordspacing}{\spaceskip=0pt\relax}
\providecommand{\BIBentryALTinterwordstretchfactor}{4}
\providecommand{\BIBentryALTinterwordspacing}{\spaceskip=\fontdimen2\font plus
\BIBentryALTinterwordstretchfactor\fontdimen3\font minus
  \fontdimen4\font\relax}
\providecommand{\BIBforeignlanguage}[2]{{%
\expandafter\ifx\csname l@#1\endcsname\relax
\typeout{** WARNING: IEEEtran.bst: No hyphenation pattern has been}%
\typeout{** loaded for the language `#1'. Using the pattern for}%
\typeout{** the default language instead.}%
\else
\language=\csname l@#1\endcsname
\fi
#2}}
\providecommand{\BIBdecl}{\relax}
\BIBdecl

\bibitem{Carofiglio-LAC}
G.~Carofiglio, L.~Mekinda, and L.~Muscariello, ``Lac: Introducing latency-aware
  caching in information-centric networks,'' in \emph{2015 IEEE 40th Conference
  on Local Computer Networks (LCN)}, Oct 2015, pp. 422--425.

\bibitem{Carofiglio-focal}
------, ``Focal: Forwarding and caching with latency awareness in
  information-centric networking,'' in \emph{2015 IEEE Globecom Workshops (GC
  Wkshps)}, Dec 2015, pp. 1--7.

\bibitem{Arald-costaware}
A.~Araldo, D.~Rossi, and F.~Martignon, ``Cost-aware caching: Caching more
  (costly items) for less (isps operational expenditures),'' \emph{IEEE
  Transactions on Parallel and Distributed Systems}, vol.~27, no.~5, pp.
  1316--1330, May 2016.

\bibitem{Thomas-objectoriented}
Y.~Thomas, G.~Xylomenos, C.~Tsilopoulos, and G.~C. Polyzos, ``Object-oriented
  packet caching for icn,'' in \emph{Proceedings of the 2Nd ACM Conference on
  Information-Centric Networking}, ser. ACM-ICN '15.\hskip 1em plus 0.5em minus
  0.4em\relax New York, NY, USA: ACM, 2015, pp. 89--98.

\bibitem{Nguyen-congestionprice}
D.~Nguyen, K.~Sugiyama, and A.~Tagami, ``Congestion price for cache management
  in information-centric networking,'' in \emph{2015 IEEE Conference on
  Computer Communications Workshops (INFOCOM WKSHPS)}, April 2015, pp.
  287--292.

\bibitem{modeling}
\BIBentryALTinterwordspacing
G.~Carofiglio, M.~Gallo, L.~Muscariello, and D.~Perino, ``Modeling data
  transfer in content-centric networking,'' in \emph{Proceedings of the 23rd
  International Teletraffic Congress}, ser. ITC '11.\hskip 1em plus 0.5em minus
  0.4em\relax International Teletraffic Congress, 2011, pp. 111--118. [Online].
  Available: \url{http://dl.acm.org/citation.cfm?id=2043468.2043487}
\BIBentrySTDinterwordspacing

\bibitem{caching}
W.~K. Chai, D.~He, I.~Psaras, and G.~Pavlou, ``Cache “less for more” in
  information-centric networks,'' in \emph{International Conference on Research
  in Networking}.\hskip 1em plus 0.5em minus 0.4em\relax Springer, 2012, pp.
  27--40.

\bibitem{ttlcachingDehgan}
M.~Dehghan, L.~Massoulie, D.~Towsley, D.~Menasche, and Y.~C. Tay, ``A utility
  optimization approach to network cache design,'' in \emph{IEEE INFOCOM 2016 -
  The 35th Annual IEEE International Conference on Computer Communications},
  April 2016, pp. 1--9.

\bibitem{agebasedcachingMing}
Z.~Ming, M.~Xu, and D.~Wang, ``Age-based cooperative caching in
  information-centric networking,'' in \emph{2014 23rd International Conference
  on Computer Communication and Networks (ICCCN)}, Aug 2014, pp. 1--8.

\bibitem{kurose}
M.~Badov, A.~Seetharam, J.~Kurose, V.~Firoiu, and S.~Nanda, ``Congestion-aware
  caching and search in information-centric networks,'' in \emph{Proceedings of
  the 1st international conference on Information-centric networking}.\hskip
  1em plus 0.5em minus 0.4em\relax ACM, 2014, pp. 37--46.

\bibitem{stratis}
S.~Ioannidis and E.~Yeh, ``Adaptive caching networks with optimality
  guarantees,'' in \emph{Proceedings of the 2016 ACM SIGMETRICS International
  Conference on Measurement and Modeling of Computer Science}.\hskip 1em plus
  0.5em minus 0.4em\relax ACM, 2016, pp. 113--124.

\bibitem{giovanna}
G.~Carofiglio, M.~Gallo, L.~Muscariello, M.~Papalini, and S.~Wang, ``Optimal
  multipath congestion control and request forwarding in information-centric
  networks,'' in \emph{Network Protocols (ICNP), 2013 21st IEEE International
  Conference on}, Oct 2013, pp. 1--10.

\bibitem{Posch-SAF}
D.~Posch, B.~Rainer, and H.~Hellwagner, ``Saf: Stochastic adaptive forwarding
  in named data networking,'' \emph{IEEE/ACM Transactions on Networking},
  vol.~25, no.~2, pp. 1089--1102, April 2017.

\bibitem{Detti-modeling}
A.~Detti, C.~Pisa, and N.~B. Melazzi, ``Modeling multipath forwarding
  strategies in information centric networks,'' in \emph{2015 IEEE Conference
  on Computer Communications Workshops (INFOCOM WKSHPS)}, April 2015, pp.
  324--329.

\bibitem{mircc}
M.~Mahdian, S.~Arianfar, J.~Gibson, and D.~Oran, ``Mircc: Multipath-aware icn
  rate-based congestion control,'' in \emph{Proceedings of the 3rd ACM
  Conference on Information-Centric Networking}, ser. ACM-ICN '16.\hskip 1em
  plus 0.5em minus 0.4em\relax New York, NY, USA: ACM, 2016, pp. 1--10.

\bibitem{Yi-adaptive}
C.~Yi, A.~Afanasyev, L.~Wang, B.~Zhang, and L.~Zhang, ``Adaptive forwarding in
  named data networking,'' \emph{SIGCOMM Comput. Commun. Rev.}, vol.~42, no.~3,
  pp. 62--67, Jun. 2012.

\bibitem{vip}
\BIBentryALTinterwordspacing
E.~M. Yeh, T.~Ho, M.~Burd, Y.~Cui, and D.~Leong, ``{VIP:} {A} framework for
  joint dynamic forwarding and caching in named data networks,'' \emph{CoRR},
  vol. abs/1310.5569, 2013. [Online]. Available:
  \url{http://arxiv.org/abs/1310.5569}
\BIBentrySTDinterwordspacing

\bibitem{gallager}
R.~Gallager, ``A minimum delay routing algorithm using distributed
  computation,'' \emph{Communications, IEEE Transactions on}, vol.~25, no.~1,
  pp. 73--85, Jan 1977.

\bibitem{jacobson}
V.~Jacobson, D.~K. Smetters, J.~D. Thornton, M.~F. Plass, N.~H. Briggs, and
  R.~L. Braynard, ``Networking named content,'' in \emph{Proceedings of CoNEXT
  '09}.\hskip 1em plus 0.5em minus 0.4em\relax New York, NY, USA: ACM, 2009,
  pp. 1--12.

\bibitem{yeh}
Y.~Xi and E.~Yeh, ``Node-based optimal power control, routing, and congestion
  control in wireless networks,'' \emph{Information Theory, IEEE Transactions
  on}, vol.~54, no.~9, pp. 4081--4106, Sept 2008.

\bibitem{datanetworks}
D.~Bertsekas and R.~Gallager, \emph{Data Networks (2Nd Ed.)}.\hskip 1em plus
  0.5em minus 0.4em\relax Upper Saddle River, NJ, USA: Prentice-Hall, Inc.,
  1992.

\bibitem{kelly}
F.~P. Kelly, \emph{Reversibility and stochastic networks}.\hskip 1em plus 0.5em
  minus 0.4em\relax Cambridge University Press, 2011.

\bibitem{femtocaching}
K.~Shanmugam, N.~Golrezaei, A.~G. Dimakis, A.~F. Molisch, and G.~Caire,
  ``Femtocaching: Wireless content delivery through distributed caching
  helpers,'' \emph{IEEE Transactions on Information Theory}, vol.~59, no.~12,
  pp. 8402--8413, 2013.

\bibitem{nlp}
D.~P. Bertsekas, \emph{Nonlinear programming}.\hskip 1em plus 0.5em minus
  0.4em\relax Athena scientific Belmont, 1999.

\end{thebibliography}
\end{document}